\documentclass[aps,preprintnumbers,twocolumn,showpacs,superscriptaddress,groupedaddress]{revtex4}  
\usepackage[dvips]{graphicx}
\usepackage{subfigure}
\usepackage{dcolumn}   
\usepackage{bm}        
\usepackage{amssymb}   
\usepackage[utf8]{inputenc}
\usepackage{mathrsfs}
\usepackage{amsmath}
\usepackage{amscd}
\usepackage[breaklinks=true,colorlinks=true]{hyperref}

\hypersetup{colorlinks=true,citecolor=blue,linkcolor=blue,urlcolor=blue}

\usepackage{color}

\begin{document}

\preprint{\small Ann.\ Phys.\ \textbf{437} (2022) 168739   
\ [\href{https://doi.org/10.1016/j.aop.2021.168739}{DOI: 10.1016/j.aop.2021.168739}]
}

\title{Deformations of kink tails}

\author{Petr A. Blinov}
\email{blinov.pa@phystech.edu}
\affiliation{Moscow Institute of Physics and Technology,
Dolgoprudny, Moscow Region 141700, Russia}

\author{Tatiana V. Gani}
\email{gani.t@bk.ru}
\affiliation{Faculty of Physics,
M.~V.~Lomonosov Moscow State University, Moscow 119991, Russia
}

\author{Vakhid A. Gani}
\email{vagani@mephi.ru}
\affiliation{Department of Mathematics, National Research Nuclear University MEPhI (Moscow Engineering Physics Institute),
Moscow 115409, Russia}
\affiliation{Theory Department, Institute for Theoretical and Experimental Physics of National Research Centre ``Kurchatov Institute'', 
Moscow 117218, Russia}

\begin{abstract}

We study the asymptotic properties of kinks in connection with the deformation procedure. We show that, upon deformation of the field-theoretic model, the asymptotics of kinks can change or remain unchanged, depending on the properties of the deforming function. The cases of both explicit and implicit kinks are considered. In addition, we show that the the deformation procedure can be applied to the important case of implicit kinks. We also prove that for any kink with a power-law tail, the stability potential decreases as the inverse square of the coordinate. The physical consequences of the deformation are discussed: the change of the kink mass, as well as the asymptotic behavior of the kink-antikink force.

\end{abstract}



\maketitle

\section{Introduction}
\label{sec:Introduction}

Kink solutions (kinks) are topologically nontrivial solutions of $(1+1)$-dimensional field-theoretic models with a real scalar field, the dynamics of which is given by the self-interaction (potential), leading to a nonlinear evolution equation \cite{Shnir.book.2018,Vachaspati.book.2006,Manton.book.2004,Vilenkin.book.2000,Rajaraman.book.1982}. Models that admit kink solutions have an extremely wide application in physics and arise in the description of various processes and objects. A classical example is the well-known Ginzburg--Landau theory of phase transitions based on a fourth-degree polynomial potential (the so-called $\varphi^4$ model) \cite{Ginzburg.ZhETF.1950,Landau.ZhETF.1937}. The $\varphi^4$ model has also many other applications \cite{Kevrekidis.book.2019}, for example, it is used for describing deformations in graphene \cite{Yamaletdinov.PRB.2017,Yamaletdinov.Carbon.2019}. In materials science, models with polynomial potentials of higher degrees are used to describe phase transitions and their sequences \cite{Khare.PRE.2014,Pavlov.CR.1999,Mroz.JPCM.1989,Gufan.DAN.1978}. In connection with various applications of topological solitons in condensed matter physics, see, e.g., \cite{Bishop.Physica_D.1980}. In the recent work \cite{Mendanha.CNSNS.2020}, the model with non-polynomial potential was successfully applied for describing the first-order phase transition between the lamellar and the inverse hexagonal phases in specific lipid bilayers. In various cosmological theories, domain walls separating regions of space with different properties (e.g., with different vacuum values of the Higgs field), are of great importance. In this case, a plane domain wall in the direction perpendicular to it is nothing more than a kink of the corresponding $(1+1)$-dimensional model \cite{Vachaspati.book.2006,Gani.JCAP.2018}.

Recently, kinks with power-law asymptotics has become a subject of growing interest. As a part of their study, models with polynomial potentials are considered and necessary conditions for the existence of kink solutions with power-law tails are formulated \cite{Lohe.PRD.1979,Belendryasova.CNSNS.2019,Radomskiy.JPCS.2017,Christov.PRD.2019,Christov.PRL.2019, Christov.CNSNS.2021,Manton.JPA.2019,Khare.JPA.2019,dOrnellas.JPC.2020,Campos.PLB.2021,Amado.EPJC.2020}. One of the important features of such kinks is the appearance of long-range interaction in kink-antikink and kink-kink systems. This, in turn, leads to a fundamentally new dynamics of such systems, in comparison with the case of exponential asymptotics. It should be noted that kinks with power-law asymptotics also appear in some models with non-polynomial potentials \cite{Guerrero.PLA.1998,Guerrero.Phys_A.1998,Guerrero.PRE.1997,Gomes.PRD.2012,Khare.JPA.2020}.

In this paper, we consider the asymptotic properties of kinks in connection with the deformation procedure. The deformation procedure for models with a real scalar field in $(1+1)$ dimensions was formulated by D.~Bazeia, L.~Losano, and J.M.C.~Malbouisson in 2002 \cite{Bazeia.PRD.2002} and presents a tool for building new field-theoretic models together with their kink solutions, based on a change of a field variable. The application of the deformation procedure has made it possible to obtain many new results in this area, see, e.g., \cite{Bazeia.PRD.2004,Bazeia.PRD.2006.braneworlds,Bazeia.PRD.2006,Brito.AoP.2014,Bazeia.EPJC.2018,Bazeia.EPJP.2018}. An appropriate choice of the deforming function can substantially change field-theoretic model, and hence can introduce new physics. When applying the deformation procedure using a strictly monotonic deforming function, the field-theoretic model is deformed in such a way that it is possible to control the amplitude of the kink (the distance between neighboring vacua of the model) and its spatial scale, without changing the overall topology of the configuration. In the context of condensed matter, in particular, this provides a convenient mechanism for adjusting the mass gap for fermionic states, which is in demand in applications. Moreover, the use of more complex deformation functions allows generating new models with more complicated topology, together with their kinks \cite{Brito.AoP.2014}. Another option is to deform a model which supports non-topological lump-like structures into a new model having topological kink solutions \cite{Bazeia.EPJP.2018}.

Our goal is to study in detail how the asymptotic behavior of a kink and its stability potential changes when the model is deformed. In particular, we find out how the asymptotics of kinks of the original and deformed models are related in the cases of exponential and power-law tails. We also prove that the asymptotic behavior of the stability potential is universal for all kinks having power-law tails. It should be noted that attention to the asymptotic behavior of kink solutions in connection with the calculations of interaction forces, etc., is a trend of recent years, see, e.g., \cite{Radomskiy.JPCS.2017,Christov.PRL.2019,Manton.JPA.2019,Khare.JPA.2019,dOrnellas.JPC.2020,Campos.PLB.2021,Khare.JPA.2020,Khare.PS.2019,Gani.PRD.2020.explicit}.

We also emphasize that the formalism we developed allows one to find the asymptotic behavior of the kink of a new (deformed) model even without finding the potential of the new model. Thus, we show how having in hand the potential of the original model and the deforming function, one can answer the question about the asymptotic behavior of the kink of the deformed model.

In Ref.~\cite{Bazeia.PRD.2002}, the deformation procedure was formulated for models with kink solutions known explicitly, i.e.\ in the form of the dependence $\varphi=\varphi_{\rm K}^{}(x)$. At the same time, for many models the kink solutions can be obtained only in an implicit form, $x=x_{\rm K}^{}(\varphi)$ \cite{Khare.PRE.2014,Belendryasova.CNSNS.2019,Radomskiy.JPCS.2017,Christov.PRD.2019,Christov.PRL.2019, Christov.CNSNS.2021,Manton.JPA.2019,Khare.JPA.2019,dOrnellas.JPC.2020,Campos.PLB.2021,Gani.JHEP.2015}. Notice that recently, efforts have been made to develop methods for obtaining kinks of higher-order polynomial models in an explicit form \cite{Gani.PRD.2020.explicit}; nevertheless, in many models, only implicit kinks are still available. Moreover, in some models it is even necessary to resort to numerical integration to obtain the kink solution \cite {Belendryasova.PLB.2021}. For this reason, it is important to generalize the deformation procedure to the case of models with implicit kinks. In this paper, we show how the kink of the deformed model can be found if the kink of the original model is known implicitly, that expands the ability of the deformation procedure. We illustrate all our findings (properties and methods) by appropriate examples.

As examples, we use several models that have many applications in physics. In particular, the above-mentioned $\varphi^4$ model, the $\varphi^8$ model, the sine-Gordon model, etc.\ (some related references are given below when discussing examples). As for deforming functions, we apply the hyperbolic sine and inverse trigonometric sine, which also have already been discussed in the literature in the context of deformation procedure.

We begin in Section \ref{sec:Kinks} by providing some basic facts about kink solutions in $(1+1)$-dimensional models that are used below. In Section \ref{sec:Deformation_procedure}, we describe the deformation procedure as it was originally introduced, and then we show how it can be applied to models with implicit kinks. Section \ref{sec:Change_good} is devoted to the study of how the exponential and power-law asymptotics of kinks are transformed when the model is deformed by a function with bounded derivative. The case of a deforming function with infinite derivative is discussed in Section \ref{sec:Change_not_good}. Section \ref{sec:Forces} presents some estimations and a discussion of the kink-antikink forces and their changes under deformation of the field model. Further, in Section \ref{sec:Stability_potential} we prove general property of the asymptotic behavior of the stability potentials for kinks with power-law tails. In Section \ref{sec:Examples}, we illustrate the found patterns with four examples. Finally, in Section \ref{sec:Conclusion}, we summarize our findings and outline our vision of directions for future work.


\section{Kink solutions}
\label{sec:Kinks}

Consider the Lagrangian density
\begin{equation}\label{eq:Largangian}
	\mathscr{L} = \frac{1}{2} \left( \frac{\partial\varphi}{\partial t} \right)^2 - \frac{1}{2} \left( \frac{\partial\varphi}{\partial x} \right)^2 - V(\varphi),
\end{equation}
that defines a field-theoretic model with a real scalar field $\varphi(x,t)$ in $(1+1)$-dimensional space-time. Here $V(\varphi)$ is the potential of the model, which is assumed to be bounded from below and having two or more isolated degenerate minima. We assume $V(\varphi)\ge0$, and $V(\varphi)=0$ in the minimum points which are also called {\it vacua of the model}. The Lagrangian \eqref{eq:Largangian} leads to the equation of motion
\begin{equation}\label{eq:eom}
	\frac{\partial^2\varphi}{\partial t^2}-\frac{\partial^2\varphi}{\partial x^2}+\frac{dV}{d\varphi}=0.
\end{equation}
In the static case $\varphi=\varphi(x)$, and Eq.~\eqref{eq:eom} takes the form $\displaystyle\frac{d^2\varphi}{dx^2} = \frac{dV}{d\varphi}$, which, in turn, can be reduced to the ordinary differential equation of the first order:
\begin{equation}\label{eq:eqmo_BPS}
    \frac{d\varphi}{dx} = \pm\sqrt{2V}.
\end{equation}

We define {\it kink} $\varphi_{\rm K}^{}(x)$ as a monotonic function of the coordinate which is a topologically non-trivial solution of Eq.~\eqref{eq:eqmo_BPS}, satisfying the boundary conditions at spatial infinities
\begin{equation}\label{eq:asymptotics}
\lim_{x \to -\infty} \varphi_{\rm K}^{} (x) = \varphi_1^{} \quad
\mbox{and} \quad \lim_{x \to +\infty} \varphi_{\rm K}^{} (x) = \varphi_2^{},
\end{equation}
where $\varphi_1^{}$ and $\varphi_2^{}$ are two neighboring vacua of the model. Obviously, the ``$+$'' sign in the right-hand side of Eq.~\eqref{eq:eqmo_BPS} leads to the monotonically increasing function $\varphi_{\rm K}^{}(x)$ (called kink) with $\varphi_1^{}<\varphi_2^{}$ in Eq.~\eqref{eq:asymptotics}. The ``$-$'' sign leads to the so-called {\it antikink} --- a monotonically decreasing function with $\varphi_1^{}>\varphi_2^{}$ in Eq.~\eqref{eq:asymptotics}. Recall that kink (antikink) is a configuration with the lowest possible energy under the given boundary conditions \eqref{eq:asymptotics}. In other words, a kink (antikink) has the minimal energy among all possible configurations in a given {\it topological sector}, i.e.\ among all configurations that satisfy the boundary conditions \eqref{eq:asymptotics}. The kink's mass, i.e., the energy of the static kink, is 
\begin{equation}\label{eq:mass_original}
    M_{\rm K}^{} =
    \begin{cases}
    \thinspace\displaystyle\int\limits_{-\infty}^{+\infty} dx \left(\frac{d\varphi_{\rm K}^{}}{dx}\right)^2 \ \ \mbox{for explicit kink},\\ \thinspace\thinspace\displaystyle\int\limits_{\varphi_1^{}}^{\varphi_2^{}} d\varphi \left(\frac{dx_{\rm K}^{}}{d\varphi}\right)^{-1} \ \ \mbox{for implicit kink}.
    \end{cases}
\end{equation}

To conclude this section, we make some additional remarks regarding the terms and symbols used below in this paper. In Eq.~\eqref{eq:asymptotics}, two adjacent minima of the potential (vacua of the model) are denoted by $\varphi_1^{}$ and $\varphi_2^{}$. We continue using this notation in the next section describing the deformation procedure. Further, starting from Section \ref{sec:Change_good}, we use only one vacuum of the two that the kink connects. We denote this vacuum by the subscript that coincides with the superscript of the corresponding potential: $\varphi_0^{}$ is the minimum of the potential $V^{(0)}(\varphi)$, and $\varphi_1^{}$ is the minimum of the potential $V^{(1)}(\varphi)$. In addition, kink solutions, i.e.\ functions $\varphi_{\rm K}^{}(x)$ for explicit kinks and $x_{\rm K}^{}(\varphi)$ for implicit kinks, are supplied by superscripts denoting which potential the kinks correspond to. For example $\varphi_{\rm K}^{(0)}(x)$ (or $x_{\rm K}^{(0)}(\varphi)$) is a kink of model with potential $V^{(0)}(\varphi)$, $\varphi_{\rm K}^{(1)}(x)$ (or $x_{\rm K}^{(1)}(\varphi)$) is a kink of model with potential $V^{(1)}(\varphi)$, and so forth.

Finally, note that throughout this paper, without loss of generality, we consider only monotonically increasing kink solutions, i.e.\ kinks, not antikinks.


\section{Deformation procedure}
\label{sec:Deformation_procedure}

Let us briefly recall the essence of the deformation procedure \cite{Bazeia.PRD.2002}. Let there be a model with the potential $V^{(0)}(\varphi)$ and known kink solution $\varphi_{\rm K}^{(0)}(x)$ connecting the vacua $\varphi_1^{}$ and $\varphi_2^{}$ of this model. Let there also be a function $f(\varphi)$ strictly monotonically increasing on the segment $[f^{-1}(\varphi_1),f^{-1}(\varphi_2)]$. Then we can introduce new model with potential
\begin{equation}\label{eq:f-deformation}
    V^{(1)}(\varphi) = \frac{V^{(0)}[f(\varphi)]}{[f^\prime(\varphi)]^2}.
\end{equation}
In the new model, the kink connecting vacua $f^{-1}(\varphi_1^{})$ and $f^{-1}(\varphi_2^{})$ is immediately known:
\begin{equation}\label{eq:new_model_kink1}
    \varphi_{\rm K}^{(1)}(x) = f^{-1}[\varphi_{\rm K}^{(0)}(x)].
\end{equation}
The model with potential \eqref{eq:f-deformation} is called $f$-deformed (or simply deformed). The mass of the deformed kink can be obtained from the first line of Eq.~\eqref{eq:mass_original}:
\begin{equation}\label{eq:mass_deformed_explicit}
    M_{\rm K}^{(1)} =
    \int\limits_{-\infty}^{+\infty}
    \left(\frac{d\varphi_{\rm K}^{(0)}}{dx}\right)^2 \left(\frac{df^{-1}}{d\varphi}\right)^2 dx.
\end{equation}

\begin{figure*}[t!]
    \centering\subfigure[\ the case of explicit kinks]{
    \includegraphics[width=0.9
 \textwidth]{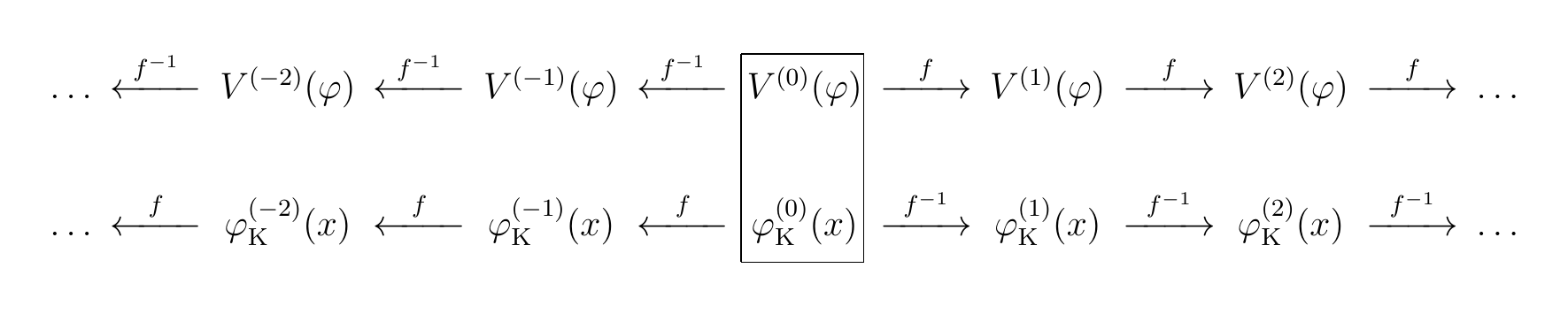}
 }
    \centering\subfigure[\ the case of implicit kinks]{
    \includegraphics[width=0.9
 \textwidth]{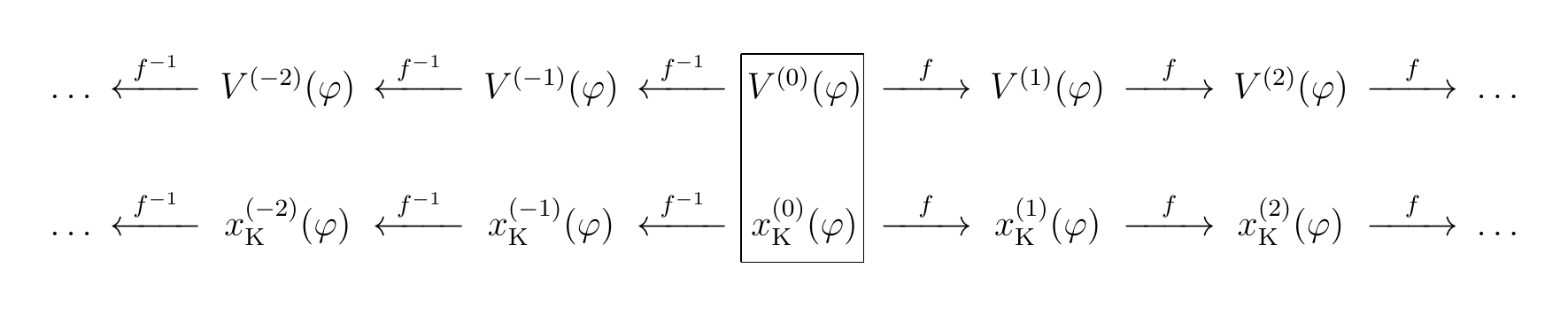}
 }
 \caption{Set of models and their kink solutions, generated by the model $V^{(0)}(\varphi)$ and the deforming function $f(\varphi)$.}
    \label{fig:set}
\end{figure*}

Of course, the same deforming function can be applied to the $f$-deformed model producing the model with potential $V^{(2)}(\varphi)$ and kink $\varphi_{\rm K}^{(2)}(x)$ obtained from $V^{(1)}(\varphi)$ and $\varphi_{\rm K}^{(1)}(x)$ using the same formulas as Eqs.~\eqref{eq:f-deformation} and \eqref{eq:new_model_kink1}.
And so one can go on and on. Moreover, we can start from the model with the potential $V^{(0)}(\varphi)$ and ``move in the opposite direction'', i.e.\ apply $f^{-1}(\varphi)$ as the deforming function. Then we get $f^{-1}$-deformed model with potential $V^{(-1)}(\varphi)$ and kink $\varphi_{\rm K}^{(-1)}(x)$, and so on. Thus, a given field-theoretic model with potential $V^{(0)}(\varphi)$ and its kink $\varphi_{\rm K}^{(0)}(x)$, together with the deforming function $f(\varphi)$, generate a countable set of new models with known kinks, see scheme in Fig.~\ref{fig:set}(a).

It so happened that in many models being used now, kink solutions can be obtained only in the implicit form, $x=x_{\rm K}^{}(\varphi)$. The deformation procedure can be successfully applied in this case.

Assume that we start again with a model with potential $V^{(0)}(\varphi)$ and (now implicit) kink $x=x_{\rm K}^{(0)}(\varphi)$. It can be shown that if we apply the deforming function $f(\varphi)$ and obtain the potential $V^{(1)}(\varphi)$ by Eq.~\eqref{eq:f-deformation}, then the implicit kink of the new model can be immediately found as $x=x_{\rm K}^{(1)}(\varphi)$, where
\begin{equation}\label{eq:first_deform_implicit_kink}
x_{\rm K}^{(1)}(\varphi)=x_{\rm K}^{(0)}(f(\varphi)).
\end{equation}
The mass of the deformed kink can be obtained from the second line of Eq.~\eqref{eq:mass_original}:
\begin{equation}\label{eq:mass_deformed_implicit}
    M_{\rm K}^{(1)} = \int\limits_{f^{-1}(\varphi_1^{})}^{f^{-1}(\varphi_2^{})} d\varphi \left(\frac{dx_{\rm K}^{(0)}(f)}{df}\right)^{-1} \left(\frac{df}{d\varphi}\right)^{-1}.
\end{equation}
Surely, as in the case of explicit kinks, deformations with deforming functions $f$ and $f^{-1}$ can be applied many times, see scheme in Fig.~\ref{fig:set}(b).

\begin{figure}[t!]
\centering
\begin{minipage}{0.8\linewidth}
\subfigure[\:Potentials]{
\includegraphics[width=\linewidth]{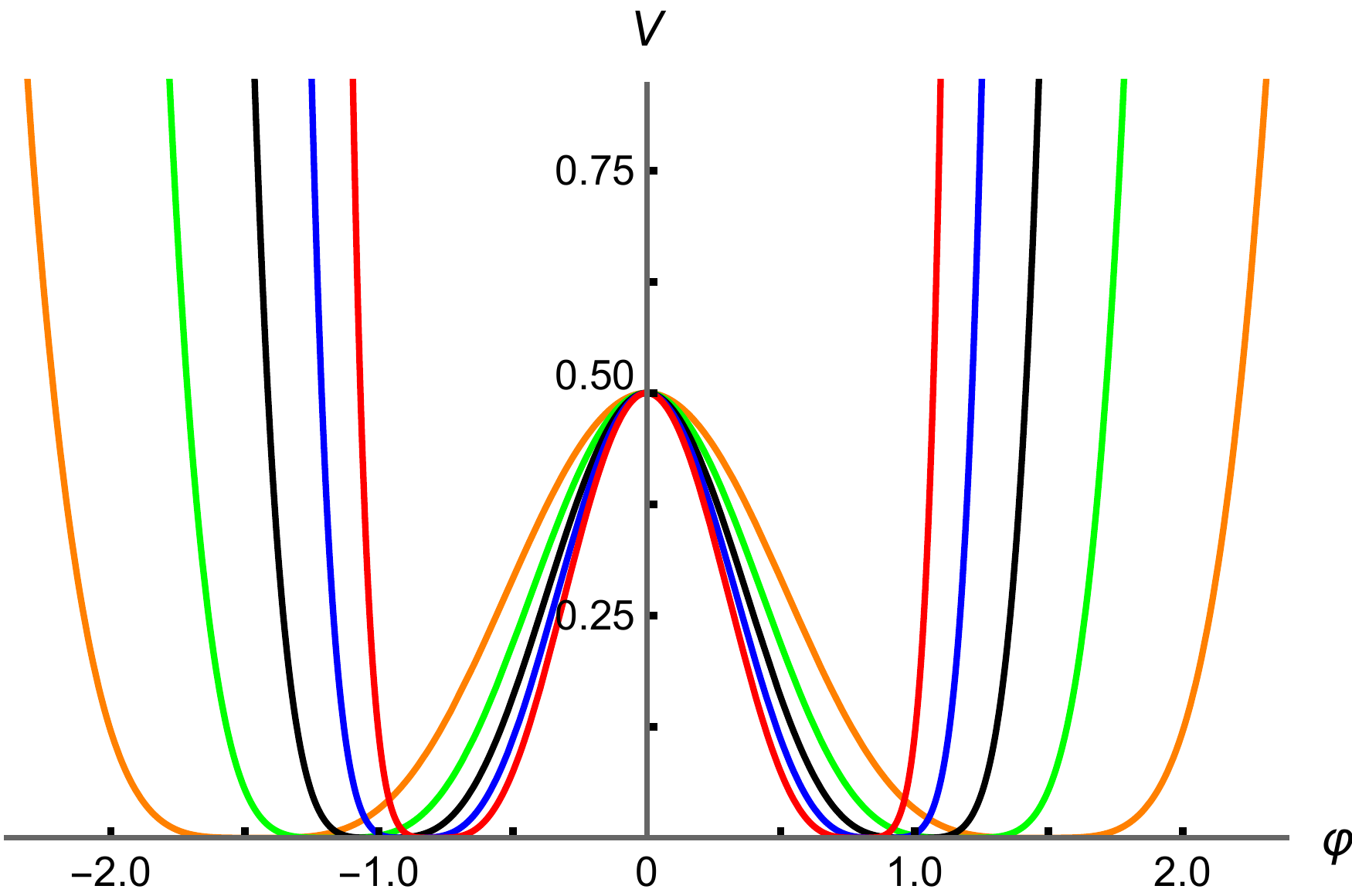}
\label{fig:many_potentials}
}
\end{minipage}
\\
\begin{minipage}{0.8\linewidth}
\subfigure[\:Kinks]{
\includegraphics[width=\linewidth]{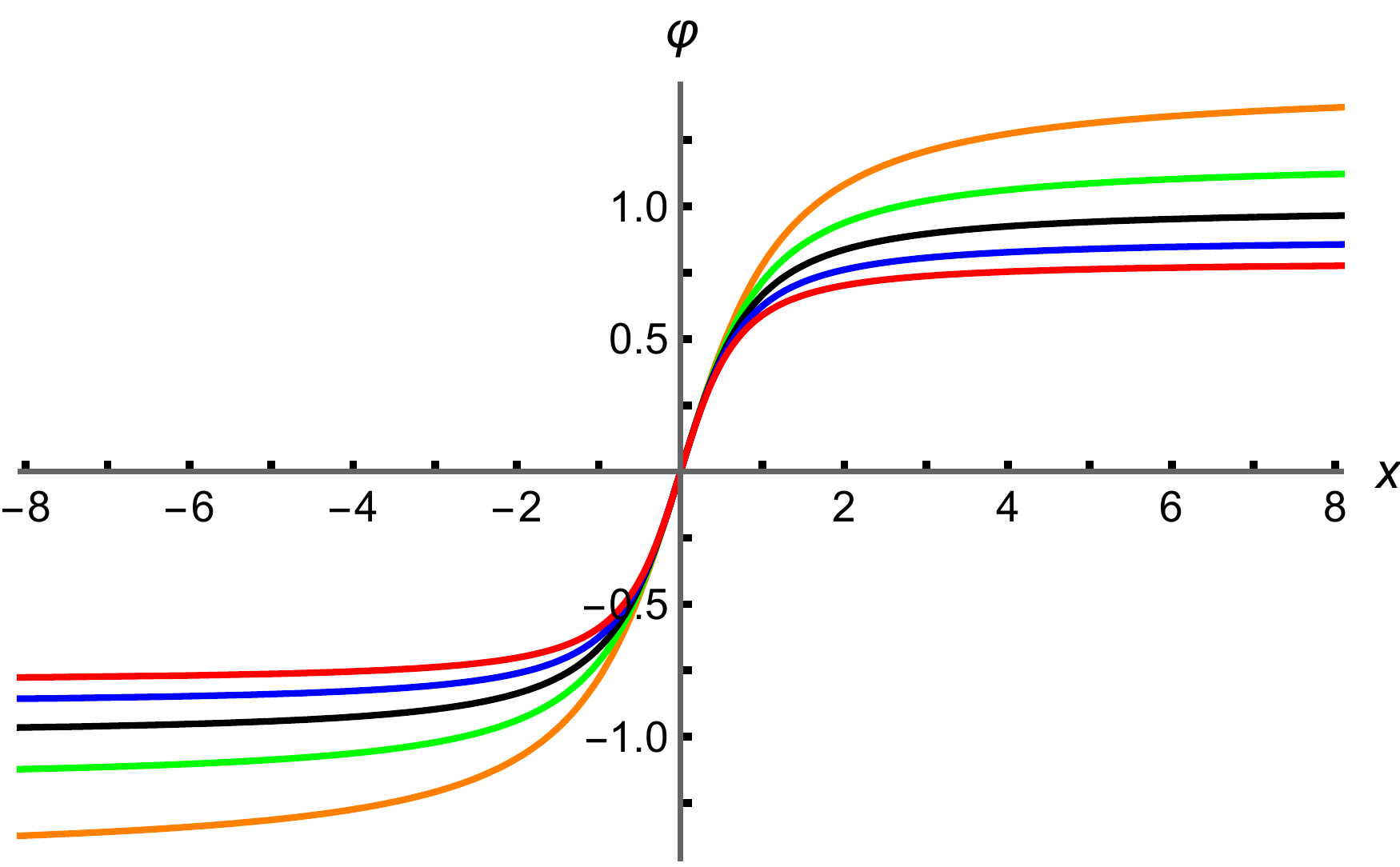}
\label{fig:many_kinks}
}
\end{minipage}
\caption{Potentials and kinks of the model \eqref{eq:potential_for_multiple_deformations} (black curves), its two successive sinh-deformations (blue and red curves), and its two successive arsinh-deformations (green and orange curves).}
\label{fig:many_potentials_and_kinks}
\end{figure}

Figure \ref{fig:many_potentials_and_kinks} shows several potentials and kinks of the sequence of models obtained from the model with the eighth degree polynomial potential
\begin{equation}\label{eq:potential_for_multiple_deformations}
    V^{(0)}(\varphi) = \frac{1}{2}\left(1-\varphi^2\right)^4
\end{equation}
using the deforming function $f(\varphi)=\sinh\varphi$. Namely, the potential and kink of the model \eqref{eq:potential_for_multiple_deformations} (black curves) and its successive deformations by the functions $f$ and $f^{-1}$ are shown. Note that kink of the model \eqref{eq:potential_for_multiple_deformations} can be found in the implicit form:
\begin{equation}\label{eq:kink_for_multiple_deformations}
        x = \underbrace{\frac{1}{2}\frac{\varphi}{1-\varphi^2} + \frac{1}{4}\ln\frac{1+\varphi}{1-\varphi}}_{x_{\rm K}^{(0)}(\varphi)}.
\end{equation}
%
%
%
%


\section{Kink's asymptotics and deformation procedure}
\label{sec:Change_good}

In this section, we will show how the asymptotics of kink changes when the model is deformed. First, for simplicity of presentation, we apply deformation to the model with polynomial potential. Then, at the end of this section, we show that the obtained transformational properties of the kink asymptotics are generalized to the case of non-polynomial potentials.

Suppose we have a model with a polynomial potential that can be represented as
\begin{equation}\label{eq:polynomial_potential_0}
    V^{(0)}(\varphi) = \frac{1}{2}\left(\varphi_0^{}-\varphi\right)^{2k} v(\varphi),
\end{equation}
where $k$ is a positive integer, and $v(\varphi)$ is some non-negative polynomial of even degree with at least one more zero at $\varphi<\varphi_0^{}$ and satisfying the condition $v(\varphi_0^{})>0$. We will consider the kink $\varphi_{\rm K}^{(0)}(x)$, interpolating between two adjacent minima of the potential \eqref{eq:polynomial_potential_0} with $\lim\limits_{x\to+\infty}\varphi_{\rm K}^{(0)}(x)=\varphi_0^{}$. For large positive $x$ the field $\varphi$ tends to the vacuum value $\varphi_0^{}$, and in Eq.~\eqref{eq:polynomial_potential_0} we set $v(\varphi)\approx v(\varphi_0^{})$. Then the potential \eqref{eq:polynomial_potential_0} can be rewritten as
\begin{equation}\label{eq:polynomial_potential_0_approx}
    V^{(0)}(\varphi) \approx \frac{1}{2}\left(\varphi_0^{}-\varphi\right)^{2k} v(\varphi_0^{}).
\end{equation}

Now we apply the deforming function $f(\varphi)$ to the potential \eqref{eq:polynomial_potential_0}. We obtain the potential $V^{(1)}(\varphi)$, which in accordance with Eq.~\eqref{eq:f-deformation}, looks like
\begin{equation}\label{eq:potential_1}
    V^{(1)}(\varphi) = \frac{\frac{1}{2}\left[\varphi_0^{}-f(\varphi)\right]^{2k}v\left(f(\varphi)\right)}{\left[f^\prime(\varphi)\right]^2}.
\end{equation}
Kink $\varphi_{\rm K}^{(1)}(x)$ of the deformed model satisfies the condition $\lim\limits_{x\to+\infty}\varphi_{\rm K}^{(1)}(x)=f^{-1}(\varphi_0^{})=\varphi_1^{}$. Our aim is to find the asymptotic behavior of $\varphi_{\rm K}^{(1)}(x)$ at large positive $x$, using the potential $V^{(0)}(\varphi)$ and deforming function $f(\varphi)$ as initial data. For $x\to+\infty$ we have $\varphi\to\varphi_1^{}$, hence in Eq.~\eqref{eq:potential_1} we replace slowly varying $v(f(\varphi))$ with $v(f(\varphi_1^{}))=v(\varphi_0^{})$, and $\left[f^\prime(\varphi)\right]^2$ with $\left[f^\prime(\varphi_1^{})\right]^2$, while we expand the difference $\varphi_0^{}-f(\varphi)$ in the vicinity of $\varphi=\varphi_1^{}$ to first order, $\varphi_0^{} - f(\varphi) \approx f^\prime(\varphi_1^{})(\varphi_1^{}-\varphi)$. As a result, for the potential $V^{(1)}(\varphi)$ at $\varphi\approx\varphi_1^{}$ we get the following approximate expression:
\begin{equation}\label{eq:polynomial_potential_1}
    V^{(1)}(\varphi) \approx \frac{1}{2} \left(\varphi_1^{}-\varphi\right)^{2k} v(\varphi_0^{}) \left[f^\prime(\varphi_1^{})\right]^{2k-2}.
\end{equation}
The structure of the potential \eqref{eq:polynomial_potential_1} obtained from \eqref{eq:potential_1} at $\varphi\approx\varphi_1^{}$ is similar to the structure of the potential \eqref{eq:polynomial_potential_0_approx} obtained from \eqref{eq:polynomial_potential_0} for $\varphi\approx\varphi_0^{}$.

We now proceed to discussion of the asymptotics of the kinks $\varphi_{\rm K}^{(0)}(x)$ and $\varphi_{\rm K}^{(1)}(x)$ of these two models, connected by the deforming function $f(\varphi)$. Because $V^{(0)}(\varphi) \sim \left(\varphi_0^{}-\varphi\right)^{2k}$ and $V^{(1)}(\varphi) \sim \left(\varphi_1^{}-\varphi\right)^{2k}$, it follows that for $k=1$ the asymptotics of both kinks are exponential, while for $k>1$ they are power-law. Below we consider these two cases separately.


\subsection{Exponential asymptotics ($k=1$)}
\label{sec:Change_good_exponential}

So let $k=1$. Using \eqref{eq:polynomial_potential_0_approx}, from Eq.~\eqref{eq:eqmo_BPS} for $x\to+\infty$, $\varphi\to\varphi_0^{}$, after integration and simple manipulations we get
\begin{equation}\label{eq:exponential_asymptotics_0}
    \varphi_{\rm K}^{(0)}(x) \approx \varphi_0^{} - \exp\left[-\sqrt{v(\varphi_0^{})}\:x\right] \quad \mbox{at} \quad x\to+\infty,
\end{equation}
The pre-exponential factor in \eqref{eq:exponential_asymptotics_0} is connected with the choice of the kink's position and has no deep meaning.

What is the asymptotic behavior of the kink $\varphi_{\rm K}^{(1)}(x)$ of the deformed model? It is seen that for $k=1$ the potential \eqref{eq:polynomial_potential_1} of the $f$-deformed model for $\varphi\approx\varphi_1^{}$ looks similar to the potential of the original model, and the factor $\left[f^\prime(\varphi_1^{})\right]^{2k-2}$ becomes equal to one, which means exactly the same asymptotic behavior of the kink of the $f$-deformed model:
\begin{equation}\label{eq:exponential_asymptotics_1}
    \varphi_{\rm K}^{(1)}(x) \approx \varphi_1^{} - \exp\left[-\sqrt{v(\varphi_0^{})}\:x\right] \quad \mbox{at} \quad x\to+\infty.
\end{equation}

Thus, the analysis performed in this subsection has shown an important property of the deformation procedure: under deformation with a good enough deforming function (a strictly monotonic function with finite derivative $f^\prime(\varphi_1^{})$), the exponential asymptotics of the kink does not change: it remains exponential with the same coefficient in front of $x$. Now let's turn our attention to the case of power-law asymptotics.


\subsection{Power-law asymptotics ($k>1$)}
\label{sec:Change_good_power-law}

Let now $k>1$. Again using \eqref{eq:polynomial_potential_0_approx}, from Eq.~\eqref{eq:eqmo_BPS} for $x\to+\infty$, $\varphi\to\varphi_0^{}$, after integration and simple manipulations, we get asymptotic behavior of the kink $\varphi_{\rm K}^{(0)}(x)$ at large positive $x$:
\begin{equation}\label{eq:kink_asymptotics_0}
    \varphi_{\rm K}^{(0)}(x) \approx \varphi_0^{} - \frac{A_k^{(0)}}{x^{1/(k-1)}} \quad \mbox{at} \quad x\to+\infty,
\end{equation}
where the coefficient $A_k^{(0)}$ does not depend on $x$,
\begin{equation}\label{eq:A_0}
     A_k^{(0)} = \left[\left(k-1\right)\sqrt{v(\varphi_0^{})}\right]^{1/(1-k)}.
\end{equation}

As already mentioned, the structure of the potential \eqref{eq:polynomial_potential_1} of the $f$-deformed model at $\varphi\approx\varphi_1^{}$ is similar to the structure of the potential \eqref{eq:polynomial_potential_0_approx} at $\varphi\approx\varphi_0^{}$. Therefore, using the same technology as in obtaining the asymptotics \eqref{eq:kink_asymptotics_0}, we find the asymptotics of the kink $\varphi_{\rm K}^{(1)}(x)$ of the  $f$-deformed model for large positive $x$:
\begin{equation}\label{eq:kink_asymptotics_1}
    \varphi_{\rm K}^{(1)}(x) \approx \varphi_1^{} - \frac{A_k^{(1)}}{x^{1/(k-1)}} \quad \mbox{at} \quad x\to+\infty,
\end{equation}
where
\begin{equation}\label{eq:A_1}
     A_k^{(1)} = \frac{1}{f^\prime(\varphi_1^{})} \left[\left(k-1\right)
     \sqrt{v(\varphi_0^{})}\right]^{1/(1-k)}.
\end{equation}

Comparing the asymptotics \eqref{eq:kink_asymptotics_0} and \eqref{eq:kink_asymptotics_1}, wee see that in both cases the field approaches the vacuum value with increasing $x$ as $x^{1/(1-k)}$. However, the coefficients $A_k^{(0)}$ and $A_k^{(1)}$ are different:
\begin{equation}\label{eq:A_ratio}
    \frac{A_k^{(0)}}{A_k^{(1)}} = f^\prime(\varphi_1^{}).
\end{equation}

Thus, we conclude that under deformation by a function with finite $f^\prime(\varphi_1^{})$, the power of the coordinate does not change, but the coefficient $A_k^{}$ changes in accordance with Eq.~\eqref{eq:A_ratio}.

In conclusion of this section, we make two important remarks that significantly expand the range of applicability of the obtained results. First, we considered the case of the polynomial potential \eqref{eq:polynomial_potential_0}. However, all arguments also remain valid for a non-polynomial function $V^{(0)}(\varphi)$ with $\varphi_0^{}$ being a zero (and a minimum) of order $2k$. Then near $\varphi=\varphi_0^{}$
\begin{equation}\label{eq:nepolinom}
    V^{(0)}(\varphi) \approx \frac{1}{2} \left(\varphi-\varphi_0^{}\right)^{2k} v(\varphi_0^{})
\end{equation}
($v(\varphi_0^{})>0$ is a number), which coincides with \eqref{eq:polynomial_potential_0_approx}. All subsequent calculations are completely similar to those performed when deriving the formulas \eqref{eq:exponential_asymptotics_0}, \eqref{eq:exponential_asymptotics_1}, \eqref{eq:kink_asymptotics_0} and \eqref{eq:kink_asymptotics_1}, and lead to the same results. Thus, the asymptotic expressions \eqref{eq:exponential_asymptotics_0}, \eqref{eq:exponential_asymptotics_1}, \eqref{eq:kink_asymptotics_0} and \eqref{eq:kink_asymptotics_1}, as well as the ratio \eqref{eq:A_ratio}, are also valid for a non-polynomial potential which can be written as Eq.~\eqref{eq:nepolinom} near the vacuum $\varphi_0^{}$.

Second, the above consideration can be applied to the kink asymptotics at $x\to-\infty$. If we assume that in the model with polynomial potential \eqref{eq:polynomial_potential_0} (or with non-polynomial potential, which can be written as \eqref{eq:nepolinom}) we have a kink $\varphi_{\rm K}^{(0)}(x)$ satisfying $\lim\limits_{x\to-\infty}\varphi_{\rm K}^{(0)}(x)=\varphi_0^{}$, then for $k=1$ we get exponential left asymptotics of kink of the original model:
\begin{equation}\label{eq:exponential_asymptotics_0_left}
    \varphi_{\rm K}^{(0)}(x) \approx \varphi_0^{} + \exp\left[\sqrt{v(\varphi_0^{})}\:x\right] \quad \mbox{at} \quad  x\to-\infty,
\end{equation}
and exponential left asymptotics of the corresponding kink of the $f$-deformed model:
\begin{equation}\label{eq:exponential_asymptotics_1_left}
    \varphi_{\rm K}^{(1)}(x) \approx \varphi_1^{} + \exp\left[\sqrt{v(\varphi_0^{})}\:x\right] \quad \mbox{at} \quad x\to-\infty.
\end{equation}
For $k>1$ we get power-law left asymptotics of kink of the original model:
\begin{equation}\label{eq:kink_asymptotics_0_left}
    \varphi_{\rm K}^{(0)}(x) \approx \varphi_0^{} + \frac{A_k^{(0)}}{|x|^{1/(k-1)}} \quad \mbox{at} \quad x\to-\infty,
\end{equation}
and power-law left asymptotics of the corresponding kink of the $f$-deformed model:
\begin{equation}\label{eq:kink_asymptotics_1_left}
    \varphi_{\rm K}^{(1)}(x) \approx \varphi_1^{} + \frac{A_k^{(1)}}{|x|^{1/(k-1)}} \quad \mbox{at} \quad x\to-\infty,
\end{equation}
where the coefficients $A_k^{(0)}$ and $A_k^{(1)}$ are defined by Eqs.~\eqref{eq:A_0} and \eqref{eq:A_1}.

So, we got that when the model is deformed by a function with finite derivative in the vacuum $\varphi_1^{}$ of the deformed model, the exponential asymptotics of the kink remains exponential, and the power-law asymptotics remains power-law. In the exponential case, the coefficient in front of $x$ does not change. In the power-law case, the power of the coordinate does not change, but the coefficient $A_k^{}$ changes depending on $f^\prime(\varphi_1^{})$, Eq.~\eqref{eq:A_ratio}.

The question arises: are there deformations that change the asymptotic behavior of the kink more significantly?


\section{Deforming function with singular derivative}
\label{sec:Change_not_good}

Consider how the kink asymptotics changes if the model is deformed by a function with infinite derivative $f^\prime(\varphi_1^{})$. Let $f(\varphi)$ be defined and strictly increasing on the segment between two neighboring vacua of the deformed model, and
\begin{equation}\label{eq:approx_f}
    f(\varphi) \approx f(\varphi_1^{}) - B\left(\varphi_1^{}-\varphi\right)^\beta \quad \mbox{at} \quad \varphi \to \varphi_1^{} - 0,
\end{equation}
where $B>0$ and $\beta$ are constants, and $0<\beta<1$, i.e.\ $f^\prime(\varphi)$ tends to infinity at $\varphi \rightarrow \varphi_1^{}-0$:
\begin{equation}\label{eq:prime_f}
     f^\prime(\varphi) \approx \frac{\beta B}{\left(\varphi_1^{}-\varphi\right)^{1-\beta}} \quad \mbox{at} \quad \varphi \to \varphi_1^{} - 0.
\end{equation}
Assume that potential of the original model for $\varphi\approx\varphi_0^{}$ has the form \eqref{eq:polynomial_potential_0_approx}. Then the potential of the $f$-deformed model at $\varphi\approx\varphi_1^{}$ looks like
\begin{equation}\label{eq:f-deformed_potential_1}
     V^{(1)}(\varphi) \approx \frac{1}{2} \frac{B^{2k-2}v(\varphi_0^{})}{\beta^2}\left(\varphi_1^{}-\varphi\right)^{2+2\beta (k-1)}.
\end{equation}
We see that for $k=1$, i.e.\ in the case of exponential asymptotics of $\varphi_{\rm K}^{(0)}(x)$, the power $2+2\beta (k-1)$ equals two, which means exponential asymptotics of the deformed kink $\varphi_{\rm K}^{(1)}(x)$. However, the coefficient $1/\beta^2$ in Eq.~\eqref{eq:f-deformed_potential_1} leads to a change of the coefficient in front of $x$. From Eq.~\eqref{eq:f-deformed_potential_1} it is easy to obtain that the kink of the deformed model has the asymptotics
\begin{equation}\label{eq:f-deformed_kink_01}
    \varphi_{\rm K}^{(1)}(x) \approx \varphi_1^{} - \exp{\left[ -\frac{\sqrt{v(\varphi_0^{})}}{\beta}\:x\right]} \quad \mbox{at} \quad x \to +\infty.
\end{equation}

Thus, under deformation by a function with the derivative going to infinity, the exponential asymptotics remains exponential. However, the field of the deformed kink approaches the vacuum value faster.

Consider the case $k>1$. Now the power $2+2\beta (k-1)$ in Eq.~\eqref{eq:f-deformed_potential_1} is necessarily greater than two, which means power-law asymptotic behavior of the deformed kink. Calculations similar to those of Section \ref{sec:Change_good} lead to the following expression:
\begin{equation}\label{eq:f-deformed_kink_1}
    \varphi_{\rm K}^{(1)}(x) \approx \varphi_1^{} - \frac{B_k^{(1)}}{x^{\frac{1}{\beta(k-1)}}} \quad \mbox{at} \quad x \to +\infty,
\end{equation}
where
\begin{equation}
    B_k^{(1)} = \left[\left(k-1\right)B_{}^{k-1}\sqrt{v(\varphi_0^{})}\right]^{\frac{1}{\beta (1-k)}}.
\end{equation}

Thus, under deformation by a function with the derivative going to infinity, the power-law asymptotics remains power-law. However, in contrast to the case of finite $f^\prime(\varphi_1^{})$, the power of the coordinate in the denominator of Eq.~\eqref{eq:f-deformed_kink_1} increases (since $0<\beta<1$), i.e.\ the field of the deformed kink approaches the vacuum value faster.


\section{Kink-antikink interactions and deformation procedure}
\label{sec:Forces}

The interaction between a kink and an antikink placed at a large distance is determined by their asymptotic behavior. In the case of exponential tails, the force of kink-antikink interaction decreases exponentially with distance, see, e.g., \cite[Ch.~5]{Manton.book.2004}. Thus, when the deforming function has finite derivative, the exponential dependence of the interaction force does not change --- the dependence remains exponential with the same exponent.

If the deforming function has an infinite derivative, i.e.\ behaves in accordance with \eqref{eq:approx_f} and \eqref{eq:prime_f}, then the force of kink-antikink interaction of the deformed model will fall off exponentially, but faster, due to the faster approach of the field of the deformed kink to the vacuum value.

As for the case of power-law asymptotics, in Ref.~\cite{Christov.PRL.2019} it was shown that for the model $V(\varphi)=\displaystyle\frac{1}{2}(1-\varphi^2)^2\varphi^{2k}$ with $k\ge 2$, the force of the kink-antikink attraction decays with the $\displaystyle\frac{2k}{k-1}$th power of their separation, if the kink and antikink are turned towards each other by power-law tails. From the analysis of Section \ref{sec:Change_good_power-law}, we see that in the case of the deforming function with finite derivative, the power $k$ does not change, hence the power of separation in the kink-antikink force does not change too. At the same time, Eq.~\eqref{eq:A_ratio} states that, e.g., for $f^{\prime}(\varphi_1^{})>1$ the asymptotics \eqref{eq:kink_asymptotics_1} of the deformed kink approaches the vacuum $\varphi_1^{}$ faster. It is natural to assume that the force of interaction in this case will be somewhat smaller.

On the other hand, from the analysis performed in Ref.~\cite{Christov.PRL.2019} it is easy to obtain that the acceleration of kink of the deformed model will be less than the acceleration of the kink of the original model. Together with smaller mass of the deformed kink (if $f^{\prime}(\varphi)>1$ for all $\varphi$ in between the vacua of the deformed model, see Eqs.~\eqref{eq:mass_deformed_explicit} and \eqref{eq:mass_deformed_implicit}) this results in a somewhat smaller force. Of course, for $f^{\prime}(\varphi_1^{})<1$ the situation is reversed --- the kink-antikink interaction becomes somewhat stronger when the model is deformed.

If the model is deformed by a function with infinite derivative $f^{\prime}(\varphi_1^{})$, the power $2k$ in Eq.~\eqref{eq:polynomial_potential_0_approx} is converted to $2+2\beta(k-1)$ in Eq.~\eqref{eq:f-deformed_potential_1}. According to this, the force of interaction in the deformed model decays with the $\displaystyle\frac{2+2\beta(k-1)}{\beta(k-1)}$th power of the kink-antikink separation. Taking into account that $k>1$ and $0<\beta<1$, we get a faster power-law decay of force with distance.


\section{Stability potential of kink with power-law tails}
\label{sec:Stability_potential}

For many processes involving kinks, the presence or absence of a vibrational mode in the excitation spectrum of a kink is of great importance, see \cite[Sec.~2.2]{Shnir.book.2018}. The excitation spectrum of a kink is determined by the discrete spectrum of the Sturm--Liouville problem (an eigenvalue problem of the type of time-independent Schr\"odinger equation),
\begin{equation}\label{eq:Schrodinger}
    \left[-\frac{d^2}{dx^2} + U(x)\right]\eta(x) = \omega^2\eta(x),
\end{equation}
which is obtained in a linear approximation in perturbation of a static kink. The {\it stability potential} $U(x)$ (sometimes also called {\it quantum mechanical potential}) is determined by the potential of the model $V(\varphi)$:
\begin{equation}\label{eq:Schrod_potential}
    U(x) = \left.\frac{d^2V}{d\varphi^2}\right|_{\varphi_{\rm K}^{}(x)},
\end{equation}
It can be shown that the derivative of the kink  $\eta_0^{}(x)=\varphi_{\rm K}^{\prime}(x)$ is the eigenfunction corresponding to the eigenvalue $\omega_0^{}=0$, see, e.g., \cite[Sec.~4]{Belendryasova.CNSNS.2019} for more details. Then from \eqref{eq:Schrodinger} it is easy to see that
\begin{equation}
    U(x) = \frac{\varphi_{\rm K}^{\prime\prime\prime}(x)}{\varphi_{\rm K}^{\prime}(x)}.
\end{equation}
Using this expression, we immediately establish a general fact: in the case of any power-law kink's asymptotics, the stability potential behaves universally: $U(x)\sim 1/x^2$. Using the asymptotics \eqref{eq:kink_asymptotics_0}, \eqref{eq:kink_asymptotics_1}, \eqref{eq:kink_asymptotics_0_left}, and \eqref{eq:kink_asymptotics_1_left} of the kinks $\varphi_{\rm K}^{(0)}(x)$ and $\varphi_{\rm K}^{(1)}(x)$, in the case of a strictly monotonic deforming function with finite derivative, we obtain
\begin{equation}
    U^{(1)}(x) = U^{(0)}(x) \approx \frac{k(2k-1)}{(k-1)^2}\cdot\frac{1}{x^2} \quad \mbox{at} \quad x \to \pm\infty,
\end{equation}
i.e., the asymptotics of the stability potential does not change upon deformation of the kink. On the other hand, in the case of a deforming function with infinite derivative, at $x\to\pm\infty$ we obtain
\begin{equation}
    U^{(1)}(x) \approx \frac{(1+\beta(k-1))(1+2\beta(k-1))}{\beta^2(k-1)^2}\cdot\frac{1}{x^2},
\end{equation}
i.e., the coefficient in front of $1/x^2$ changes:
\begin{equation}
    U^{(1)}(x) = \frac{(1+\beta(k-1))(1+2\beta(k-1))}{\beta^2k(2k-1)} \cdot U^{(0)}(x).
\end{equation}
Note that the coefficient in the right-hand side grows as $1/\beta^2$ at $\beta\to 0$. Moreover, in all cases the potential $U(x)$ approaches zero from above as $x\to\pm\infty$.

It should be noted that understanding the asymptotic behavior of the stability potential is important for studying the excitation spectrum of a kink. Although the excitation spectrum of a kink with at least one power-law asymptotics cannot contain vibrational (normal) modes, nevertheless, it can contain quasi-normal mode(s). The question of the existence of quasi-normal modes and their relation to resonance phenomena in the kink-antikink scattering has not yet been studied. The universality of the asymptotics that we have shown, as well as its transformational properties with respect to the deformation procedure, is important for further progress in this direction.


\section{Examples}
\label{sec:Examples}

In this section, we provide several examples illustrating the application of the developed formalism. We have selected the models from the most common in various applications.

\subsection{First example}
\label{sec:Example_1}

As an example illustrating the transformation of exponential asymptotics, consider deformation of the well-known model $\varphi^4$ using the deforming function $f(\varphi)=\sinh\varphi$. Properties of kinks of the $\varphi^4$ model and its modifications have been studied very well, see, e.g., \cite{Kevrekidis.book.2019,Kudryavtsev.JETPLett.1975,Sugiyama.PTP.1979,Campbell.PhysD.1983,Belova.PhysD.1988,Anninos.PRD.1991,Goodman.SIAM_JADS.2005,Dutta.PRL.2008,Weigel.PRD.2016,Weigel.JPCS.2014,Dorey.JHEP.2017,Moradi.CNSNS.2017,Dorey.PLB.2018,Adam.JHEP.2019,Adam.PRD.2020,Yan.PLB.2020,Askari.CSF.2020,Adam.PRE.2020,Mohammadi.arXiv.2020,Alonso-Izquierdo.PRD.2021,Manton.PRD.2021}.

We take the potential $V^{(0)}(\varphi)$ in the form
\begin{equation}\label{eq:First_example_potential}
    V^{(0)}(\varphi) = \frac{1}{2}\left(1-\varphi^2\right)^2.
\end{equation}
The corresponding kink is
\begin{equation}\label{eq:First_example_kink}
    \varphi_{\rm K}^{(0)}(x)=\tanh x.
\end{equation}

\begin{figure}[t!]
\centering
\begin{minipage}{0.8\linewidth}
\subfigure[\:Potentials]{
\includegraphics[width=\linewidth]{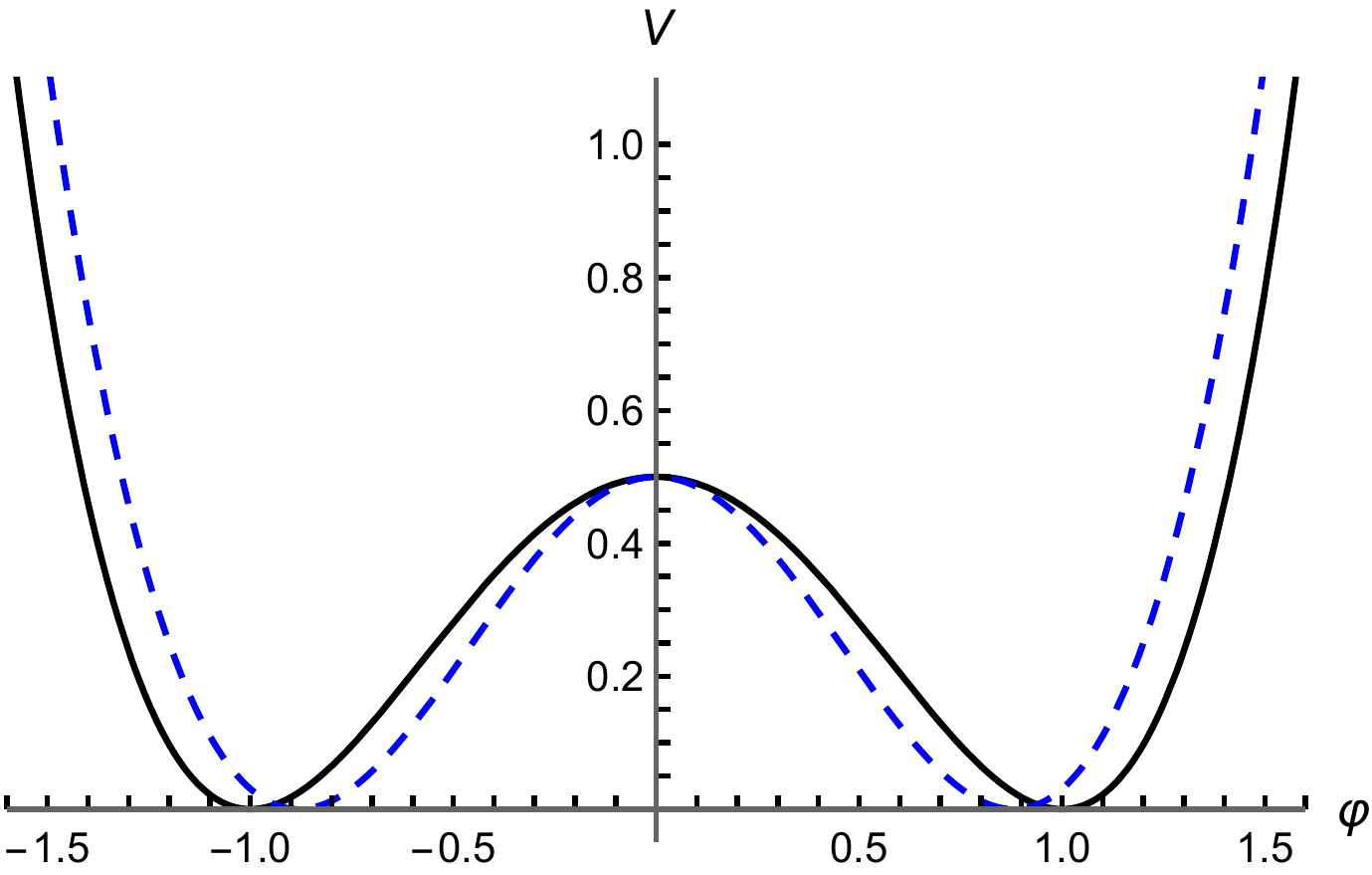}
\label{fig:potentials_example_1}
}
\end{minipage}
\\
\begin{minipage}{0.8\linewidth}
\subfigure[\:Kinks]{
\includegraphics[width=\linewidth]{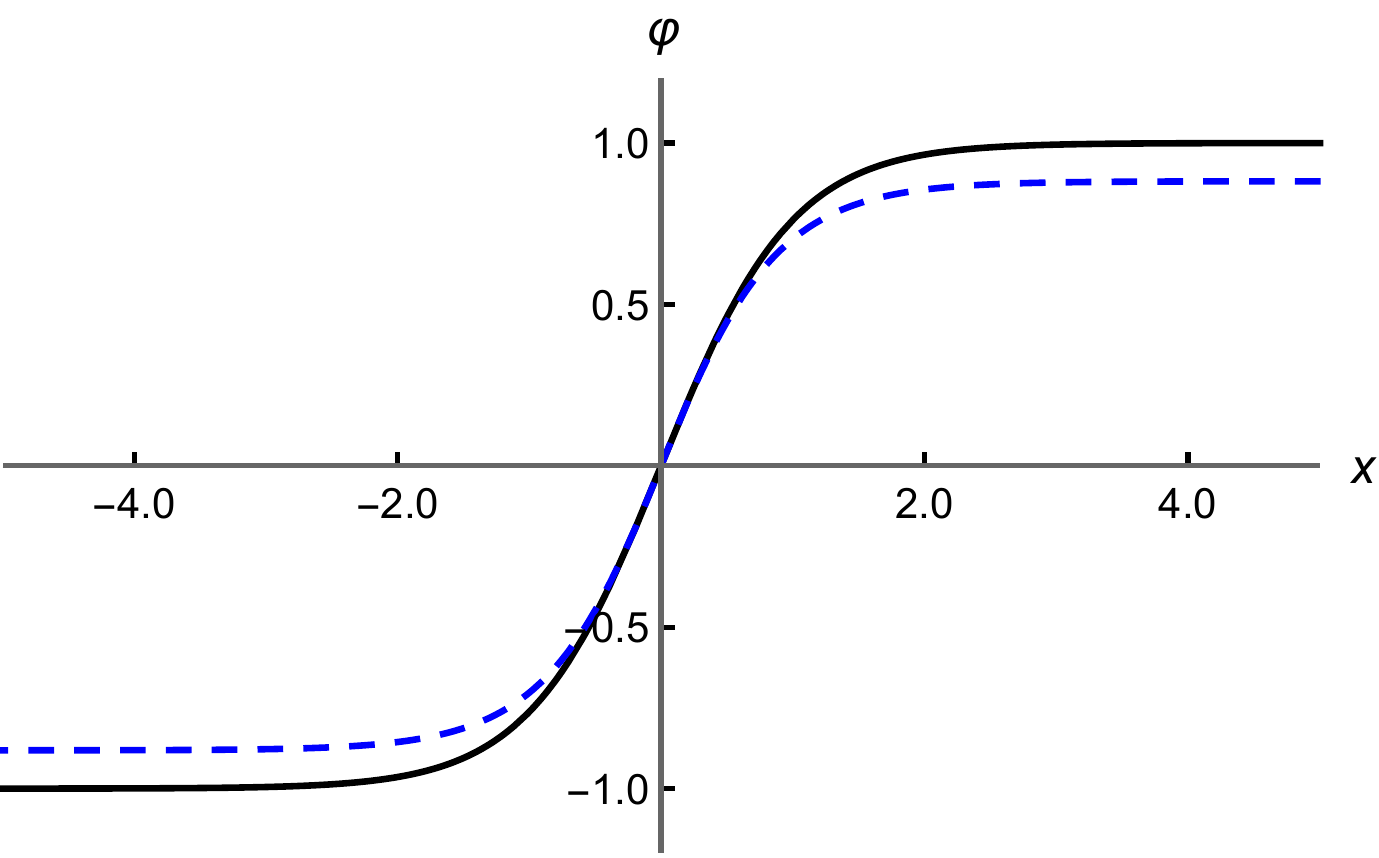}
\label{fig:kinks_example_1}
}
\end{minipage}
\caption{Potentials and kinks of the $\varphi^4$ model \eqref{eq:First_example_potential} (black solid curves) and of the sinh-deformed $\varphi^4$ model \eqref{eq:First_example_deformed_potential} (blue dashed curves).}
\label{fig:potentials_and_kinks_example_1}
\end{figure}

We deform the $\varphi^4$ model and obtain the sinh-deformed $\varphi^4$ model \cite{Bazeia.EPJC.2018}. Potential and kink of the sinh-deformed $\varphi^4$ model are
\begin{equation}\label{eq:First_example_deformed_potential}
    V^{(1)}(\varphi) = \frac{1}{2}\: \text{sech}^2\varphi\left(1-\sinh^2\varphi\right)^2
\end{equation}
and
\begin{equation}\label{eq:First_example_deformed_kink}
    \varphi_{\rm K}^{(1)}(x) = \text{arsinh}(\tanh x),
\end{equation}
respectively. Potentials and kinks of both models are shown in Fig.~\ref{fig:potentials_and_kinks_example_1}.
%
%
%
The kink \eqref{eq:First_example_deformed_kink} connects minima $\pm\mbox{arsinh}\:1$ of the potential \eqref{eq:First_example_deformed_potential}.

Now, we apply the formalism developed in Section \ref{sec:Change_good_exponential}. Eqs.~\eqref{eq:exponential_asymptotics_0} and \eqref{eq:exponential_asymptotics_0_left} give the asymptotic behavior of the $\varphi^4$ kink at $x\to+\infty$ and $x\to-\infty$:
\begin{equation}\label{eq:First_example_fully_asymptotics}
    \varphi_{\rm K}^{(0)}(x) \approx
    \begin{cases}
    -1 + e^{2x} \quad \mbox{at} \quad x\to-\infty,\\
    \thinspace 1 - e^{-2x} \quad \mbox{at} \quad x\to+\infty.
    \end{cases}
\end{equation}
To find out asymptotics of the sinh-deformed $\varphi^4$ kink, we use Eq.~\eqref{eq:exponential_asymptotics_1} with $\varphi_{1}^{}=\mbox{arsinh}\:1$ and Eq.~\eqref{eq:exponential_asymptotics_1_left} with $\varphi_{1}^{}= -\mbox{arsinh}\:1$. We obtain
\begin{equation}\label{eq:First_example_fully_deformed_asymptotics}
    \varphi_{\rm K}^{(1)}(x) \approx
    \begin{cases}
    -\mbox{arsinh}\:1 + e^{2x} \quad \mbox{at} \quad x\to-\infty,\\
    \thinspace \mbox{arsinh}\:1 - e^{-2x} \quad \mbox{at} \quad x\to+\infty.
    \end{cases}
\end{equation}
We see that the kinks of the original and deformed models exhibit the same asymptotic behavior. Of course, these asymptotics (up to pre-exponential factors) can also be obtained directly from the kinks \eqref{eq:First_example_kink} and \eqref{eq:First_example_deformed_kink}

The kink-antikink force decays exponentially in both cases (the $\varphi^4$ case is discussed, e.g., in \cite[Sec.~5.2]{Manton.book.2004}). The $\varphi^4$ kink's mass is $M_{\rm K}^{(0)}=4/3$, while for the sinh-deformed $\varphi^4$ kink we have $M_{\rm K}^{(1)}=\pi-2$. The deformed kink is lighter, $M_{\rm K}^{(1)}<M_{\rm K}^{(0)}$, according to Eq.~\eqref{eq:mass_deformed_explicit} and $f^{\prime}(\varphi)\ge 1$ for $-\mbox{arsinh}\:1\le\varphi\le\mbox{arsinh}\:1$.


\subsection{Second example}
\label{sec:Example_2}

Consider the $\varphi^8$ model with the potential
\begin{equation}\label{eq:second_potential_0}
    V^{(0)}(\varphi)=\frac{1}{2}\:\varphi^4\left(1-\varphi^2\right)^2,
\end{equation}
which is currently being actively studied \cite{Khare.PRE.2014,Lohe.PRD.1979,Belendryasova.CNSNS.2019,Radomskiy.JPCS.2017,Christov.PRD.2019,Christov.PRL.2019, Christov.CNSNS.2021,Manton.JPA.2019,Khare.JPA.2019,dOrnellas.JPC.2020,Gani.JHEP.2015}. Potential \eqref{eq:second_potential_0} has three degenerate minima: $\varphi=0$, $\varphi=\pm 1$. Consider the kink connecting $-1$ and $0$, it has exponential left and power-law right asymptotics.

\begin{figure}[t!]
\centering
\begin{minipage}{0.8\linewidth}
\subfigure[\:Potentials]{
\includegraphics[width=\linewidth]{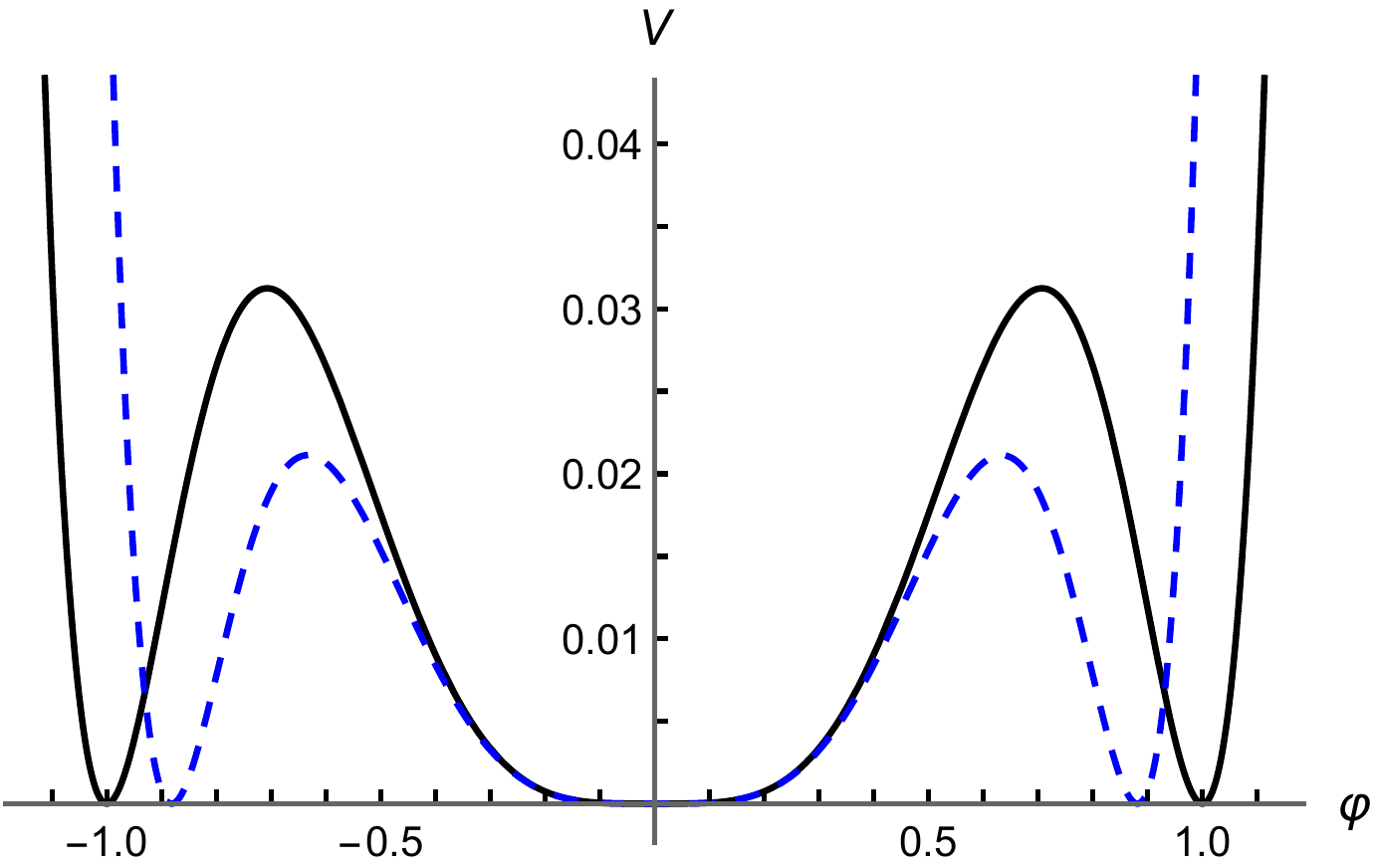}
\label{fig:potentials_example_2}
}
\end{minipage}
\\
\begin{minipage}{0.8\linewidth}
\subfigure[\:Kinks]{
\includegraphics[width=\linewidth]{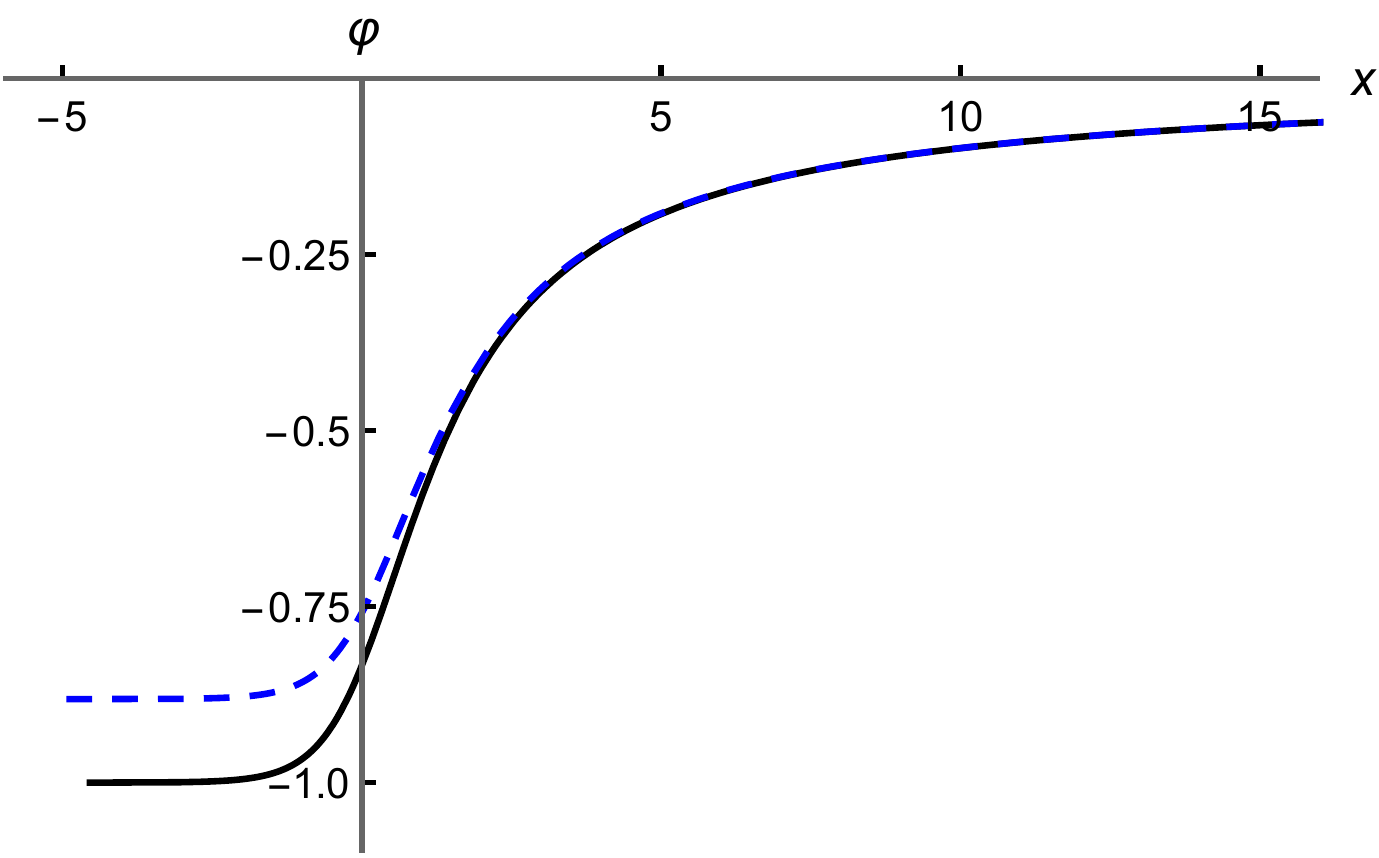}
\label{fig:kinks_example_2}
}
\end{minipage}
\caption{Potentials and kinks of the $\varphi^8$ model \eqref{eq:second_potential_0} (black solid curves) and of the sinh-deformed $\varphi^8$ model \eqref{eq:second_potential_1} (blue dashed curves).}
\label{fig:potentials_and_kinks_example_2}
\end{figure}

Without getting the kink of the model \eqref{eq:second_potential_0}, we can find its asymptotics. Then we apply the deforming function $f(\varphi)=\sinh\varphi$, and obtain the sinh-deformed $\varphi^8$ model:
\begin{equation}\label{eq:second_potential_1}
    V^{(1)}(\varphi) = \frac{1}{2}\sinh^2\varphi\left(1-\sinh^2\varphi\right)^2 \tanh^2\varphi.
\end{equation}
Potentials and kinks of both models are presented in Fig.~\ref{fig:potentials_and_kinks_example_2}.
%
%
%
Next, without getting the deformed kink, we can find its asymptotics. Besides, the kink $\varphi_{\rm K}^{(0)}(x)$ of the model \eqref{eq:second_potential_0} can only be found implicitly in the form $x=x_{\rm K}^{(0)}(\varphi)$. To obtain the kink $\varphi_{\rm K}^{(1)}(x)$ of the deformed model, we will use the technology we have developed in Section \ref{sec:Deformation_procedure}, Eq.~\eqref{eq:first_deform_implicit_kink}.

The left asymptotics of the $\varphi^8$ kink is obtained from Eq.~\eqref{eq:exponential_asymptotics_0_left}, and the right --- from Eq.~\eqref{eq:kink_asymptotics_0}:
\begin{equation}\label{eq:approx_varphi_0}
    \varphi_{\rm K}^{(0)}(x) \approx
    \begin{cases}
    -1 + e^{2x} \quad \mbox{at} \quad x \to -\infty,\\
    \thinspace 0 - \displaystyle\frac{1}{x} \quad \mbox{at} \quad x \to +\infty.
    \end{cases}
\end{equation}
Left and right asymptotics of the corresponding kink of the sinh-deformed $\varphi^8$ model are defined by Eqs.~\eqref{eq:exponential_asymptotics_1_left} and \eqref{eq:kink_asymptotics_1}, respectively:
\begin{equation}\label{eq:approx_varphi_1}
    \varphi_{\rm K}^{(1)}(x) \approx
    \begin{cases}
    -\mbox{arsinh}\: 1 + e^{2x} \quad \mbox{at} \quad x \to -\infty,\\
    \thinspace 0 - \displaystyle\frac{1}{x} \quad \mbox{at} \quad x \to +\infty.
    \end{cases}
\end{equation}

The left exponential asymptotics of the kink has not changed under deformation, in accordance with the general rule obtained in Section~\ref{sec:Change_good_exponential}. Moreover, we see that the right power-law asymptotics has not changed too, as a consequence of Eq.~\eqref{eq:A_ratio}, since in this case $f^\prime(0)=1$.

Now we use the fact that the kink of the model \eqref{eq:second_potential_0} can be found in the implicit form $x=x_{\rm K}^{(0)}(\varphi)$. Simple calculations yield
\begin{equation}\label{eq:simple_0}
    x_{\rm K}^{(0)}(\varphi)=-\frac{1}{\varphi}+\frac{1}{2}\ln{\frac{1+\varphi}{1-\varphi}}.
\end{equation}
Using this expression, one can obtain asymptotics which coincide with \eqref{eq:approx_varphi_0} up to the pre-exponential factor in the left asymptotics.

The corresponding kink of the sinh-deformed $\varphi^8$ model can be obtained from \eqref{eq:simple_0}, using Eq.~\eqref{eq:first_deform_implicit_kink} derived in Section \ref{sec:Deformation_procedure}:
\begin{equation}\label{eq:simple_1}
    x_{\rm K}^{(1)}(\varphi)=-\frac{1}{\sinh\varphi}+\frac{1}{2}\ln{\frac{1+\sinh\varphi}{1-\sinh\varphi}}.
\end{equation}
Its asymptotic behavior coincide with Eq.~\eqref{eq:approx_varphi_1} up to the pre-exponential factor.

The kink-antikink force for both models decays as fourth power of the inverse separation \cite{Christov.PRL.2019}. The kinks' masses are $M_{\rm K}^{(0)}=\displaystyle\frac{2}{15}\approx 0.13$ and $M_{\rm K}^{(1)}=\displaystyle\frac{5}{3}-\frac{\pi}{4}-2\:\mbox{arctg}\:(\sqrt{2}-1)\approx 0.096$. As in the previous example, the deformed kink is lighter, $M_{\rm K}^{(1)}<M_{\rm K}^{(0)}$, now according to Eq.~\eqref{eq:mass_deformed_implicit}.


\subsection{Third example}
\label{sec:Example_3}

As an example of deforming function that has unbounded derivative and satisfies the conditions \eqref{eq:approx_f} and \eqref{eq:prime_f}, we consider the function $f(\varphi)=\arcsin\varphi$. As an initial model, we take
\begin{equation}\label{eq:start_potential_0_example_3}
    V^{(0)}(\varphi)=\frac{1}{2}\cos^2\varphi.
\end{equation}
This potential is periodic with minima $\varphi=\pm\pi/2$, $\pm 3\pi/2$, $\pm 5\pi/2$, ... .
%
%
%
\begin{figure}[t!]
\centering
\begin{minipage}{0.8\linewidth}
\subfigure[\:Potentials]{
\includegraphics[width=\linewidth]{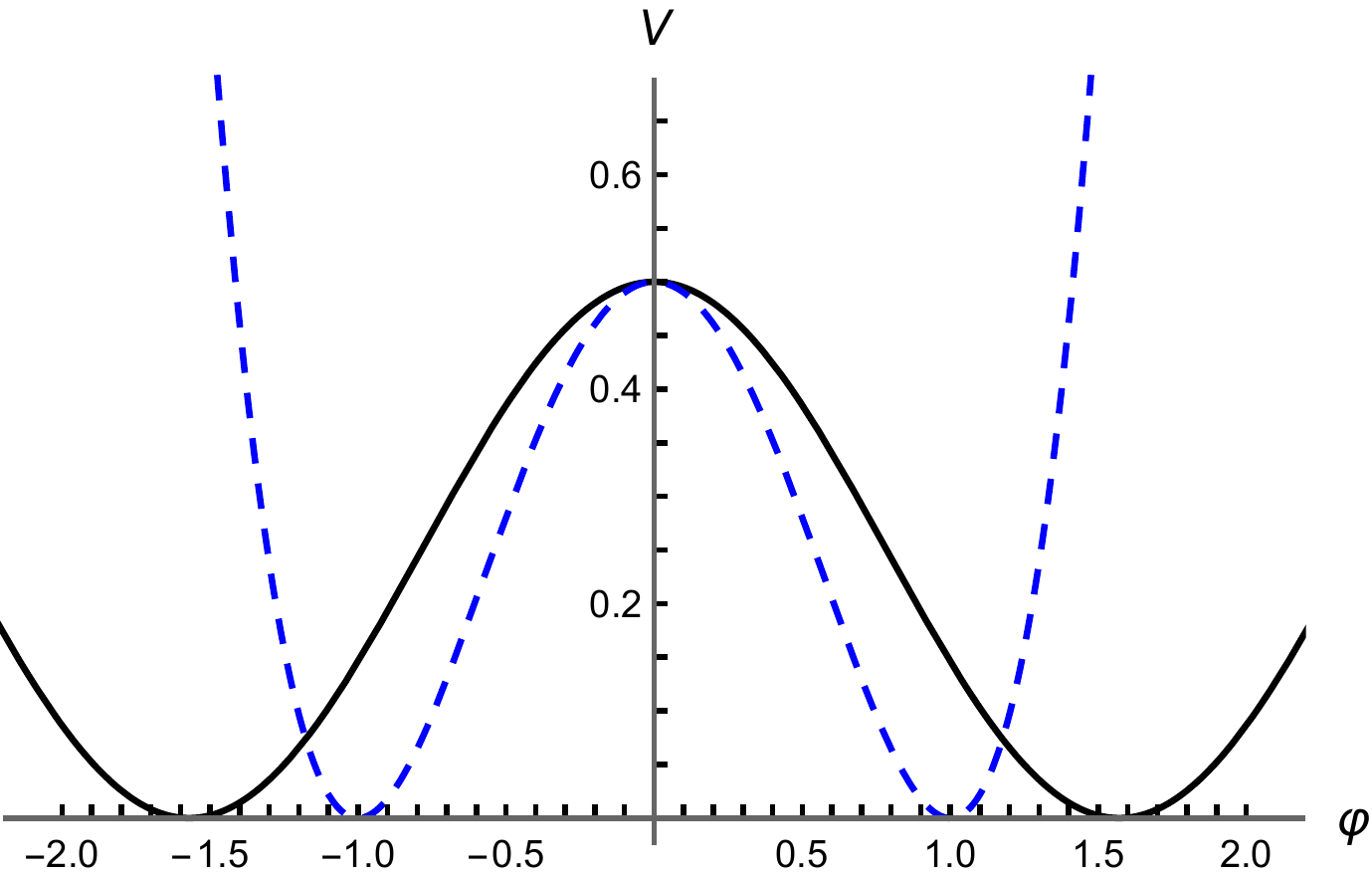}
\label{fig:potentials_example_3}
}
\end{minipage}
\\
\begin{minipage}{0.8\linewidth}
\subfigure[\:Kinks]{
\includegraphics[width=\linewidth]{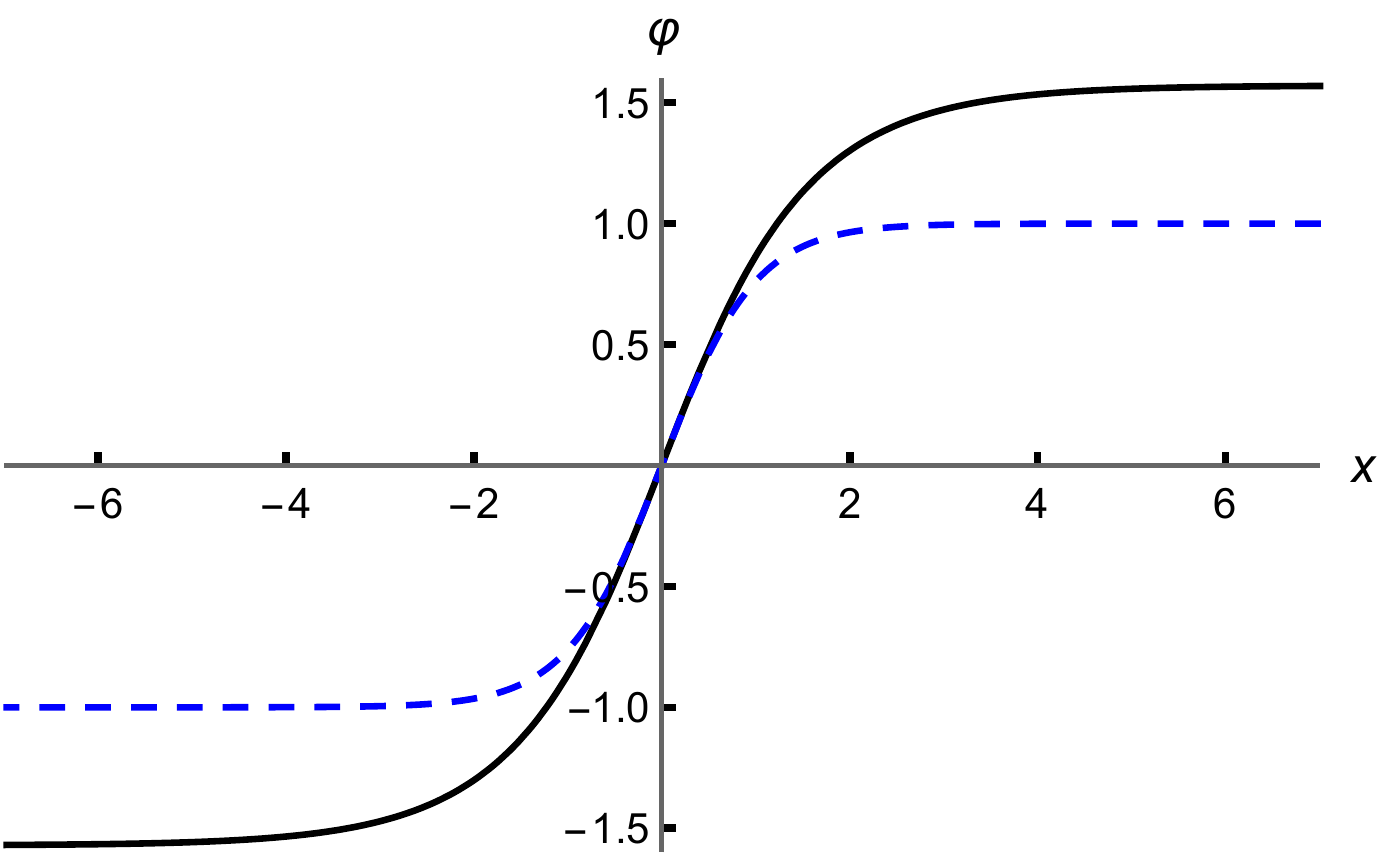}
\label{fig:kinks_example_3}
}
\end{minipage}
\caption{Potentials and kinks of the sine-Gordon model \eqref{eq:start_potential_0_example_3} (black solid curves) and of the $\varphi^4$ model \eqref{eq:new_potential_1_example_3} (blue dashed curves).}
\label{fig:potentials_and_kinks_example_3}
\end{figure}
Consider the kink connecting the vacua $-\pi/2$ and $\pi/2$:
\begin{equation}\label{eq:start_kink_0_example_3}
    \varphi_{\rm K}^{(0)}(x) = \arcsin(\tanh x).
\end{equation}
Deform the model \eqref{eq:start_potential_0_example_3} using the function $f(\varphi)=\arcsin\varphi$, and obtain the $\varphi^4$ model with potential
\begin{equation}\label{eq:new_potential_1_example_3}
V^{(1)}(\varphi) = \frac{1}{2}\left(1-\varphi^2\right)^2
\end{equation}
and kink
\begin{equation}\label{eq:new_kink_1_example_3}
    \varphi_{\rm K}^{(1)}(x) = \tanh x.
\end{equation}
Potentials and kinks of both models are shown in Fig.~\ref{fig:potentials_and_kinks_example_3}.

Asymptotics of the kink of the original model \eqref{eq:new_potential_1_example_3} can be found from Eq.~\eqref{eq:exponential_asymptotics_0}. Then, taking into account the kink's symmetry:
\begin{equation}  \label{eq:asympt_kink_s0_example_3}
    \varphi_{\rm K}^{(0)}(x) \approx
    \begin{cases}
    -\displaystyle\frac{\pi}{2} + e^{x} \quad \mbox{at} \quad x\to-\infty,\vspace{1mm}\\
    \thinspace \displaystyle\frac{\pi}{2} - e^{-x} \quad \mbox{at} \quad x\to+\infty.
    \end{cases}
\end{equation}
It can be easily seen that the asymptotics \eqref{eq:asympt_kink_s0_example_3} coincide up to the pre-exponential factors with that could be obtained from \eqref{eq:start_kink_0_example_3}. As for kink of the deformed model, its asymptotics were found and discussed in Section \ref{sec:Example_1}, see Eq.~\eqref{eq:First_example_fully_asymptotics}.

Now let's see what the formulas obtained in Section \ref{sec:Change_not_good} give. Near the vacuum $\varphi_0^{}=\pi/2$ the potential $V^{(0)}(\varphi)$ looks like $V^{(0)}(\varphi) \approx \displaystyle\frac{1}{2}\left(\frac{\pi}{2}-\varphi\right)^2$, i.e., $v(\varphi_0^{})=1$. On the other hand, for the deforming function $f(\varphi)=\arcsin\varphi$ at $\varphi\to 1-0$ we have $f(\varphi)\approx \displaystyle\frac{\pi}{2}-\sqrt{2}\: \sqrt{1-\varphi}$ and $f^\prime(\varphi) \approx \displaystyle\frac{1}{\sqrt{2}}\: \frac{1}{\sqrt{1-\varphi}}$, i.e., $B=\sqrt{2}$, $\beta=1/2$, $\varphi_1^{}=1$. Then Eq.~\eqref{eq:f-deformed_potential_1} allows us to find the behavior of the potential $V^{(1)}(\varphi)$ near the vacuum $\varphi_1^{}$: $V^{(1)}(\varphi) \approx \displaystyle\frac{1}{2}\left(1-\varphi\right)^2\cdot 4$, which, obviously, completely coincides with the behavior of the potential \eqref{eq:new_potential_1_example_3} of the $\varphi^4$ model.

Let us now see what our formulas give for asymptotics of the kink $\varphi_{\rm K}^{(1)}(x)$, i.e.\ find asymptotics of $\varphi_{\rm K}^{(1)}(x)$, using only behavior of the potential $V^{(0)}(\varphi)$ near the vacuum $\varphi_0^{}=\pi/2$ and behavior of the deforming function near the vacuum $\varphi_1^{}=1$. Then asymptotics of the deformed kink is given by Eq.~\eqref{eq:f-deformed_kink_01}. Taking into account the kink's symmetry, we obtain for $\varphi_{\rm K}^{(1)}(x)$ the same asymptotics as in the right-hand side of Eq.~\eqref{eq:First_example_fully_asymptotics}, i.e., the correct asymptotic behavior of the $\varphi^4$ kink.

Thus, this example demonstrates the change in the exponential asymptotic behavior of kink under deformation by a function that has infinite derivative. In this case, the kink of the deformed model demonstrates a faster approach of the field to vacuum values with distance from the kink localization. The exponential asymptotics remains exponential, but the coefficient in front of $x$ increases by factor $1/\beta=2$.

Note that the original model with potential \eqref{eq:start_potential_0_example_3} had a non-polynomial potential and is integrable \cite[Sec.~1.3]{Shnir.book.2018} (the well-known sine-Gordon model \cite[Sec.~5.3]{Manton.book.2004}). At the same time, after deformation by the function $f(\varphi)=\arcsin\varphi$, another well-known model (the $\varphi^4$ model) is obtained, which has polynomial potential and is no longer integrable.

The kink-antikink force for both models decays exponentially, according to exponential tails of the kinks \cite[Sec.~5]{Manton.book.2004}. The kinks' masses are $M_{\rm K}^{(0)}=2$ and $M_{\rm K}^{(1)}=4/3$, $M_{\rm K}^{(1)}<M_{\rm K}^{(0)}$, because $f^{\prime}(\varphi)\ge 1$ for $-1\le\varphi\le 1$ in Eq.~\eqref{eq:mass_deformed_explicit}.


\subsection{Fourth example}
\label{sec:Example_4}

Let's see how our formulas work in the case of power-law kink asymptotics and the deforming function with infinite derivative. As an example, consider a model with non-polynomial periodic potential
\begin{equation}\label{eq:Fourth_example_potential}
    V^{(0)}(\varphi) = \frac{1}{2}\cos^4\varphi.
\end{equation}
We take kink interpolating between the minima $-\pi/2$ and $\pi/2$. This kink is obviously symmetric.

Again, take the deforming function $f(\varphi)=\arcsin\varphi$. The vacua of the deformed model in this case are $f^{-1}(-\pi/2)=-1$ and $f^{-1}(\pi/2)=1$. The derivative of the deforming function goes to infinity at both points, and in the same way. Based on the form of the potential \eqref{eq:Fourth_example_potential} and the deforming function, we can obtain the asymptotic behavior of the kinks of the original and deformed models.

The kink $\varphi_{\rm K}^{(0)}(x)$ has power-law right asymptotics, which is given by Eq.~\eqref{eq:kink_asymptotics_0}, while its left asymptotics can be obtained from Eq.~\eqref{eq:kink_asymptotics_0_left} or by symmetry:
\begin{equation}\label{eq:Fourth_example_left_asymptotics}
    \varphi_{\rm K}^{(0)}(x) \approx
    \begin{cases}
    - \displaystyle\frac{\pi}{2} + \frac{1}{|x|} \quad \mbox{at} \quad x\to-\infty,\vspace{1mm}\\
    \thinspace \thinspace \thinspace \thinspace   \displaystyle\frac{\pi}{2} - \frac{1}{x} \quad \mbox{at} \quad x\to+\infty.
    \end{cases}
\end{equation}
The right asymptotics of the deformed kink connecting the vacua $-1$ and $1$ can be found from Eq.~\eqref{eq:f-deformed_kink_1}, its left asymptotics is symmetrical to the right:
\begin{equation}\label{eq:Fourth_example_right_asymptotics}
    \varphi_{\rm K}^{(1)}(x) \approx
    \begin{cases}
    -1 + \displaystyle\frac{1}{2x^2} \quad \mbox{at} \quad x\to-\infty,\vspace{1mm}\\
    \thinspace \thinspace \thinspace \thinspace  1 - \displaystyle\frac{1}{2x^2} \quad \mbox{at} \quad x\to+\infty.
    \end{cases}
\end{equation}

Comparing Eqs.~\eqref{eq:Fourth_example_left_asymptotics} and \eqref{eq:Fourth_example_right_asymptotics}, we see that the kink of the deformed model demonstrates a faster approach of the field to the vacuum values. Nevertheless, the power-law asymptotics remains power-law. In general, this situation is similar to the exponential case (previous example).

\begin{figure}[t!]
\centering
\begin{minipage}{0.8\linewidth}
\subfigure[\:Potentials]{
\includegraphics[width=\linewidth]{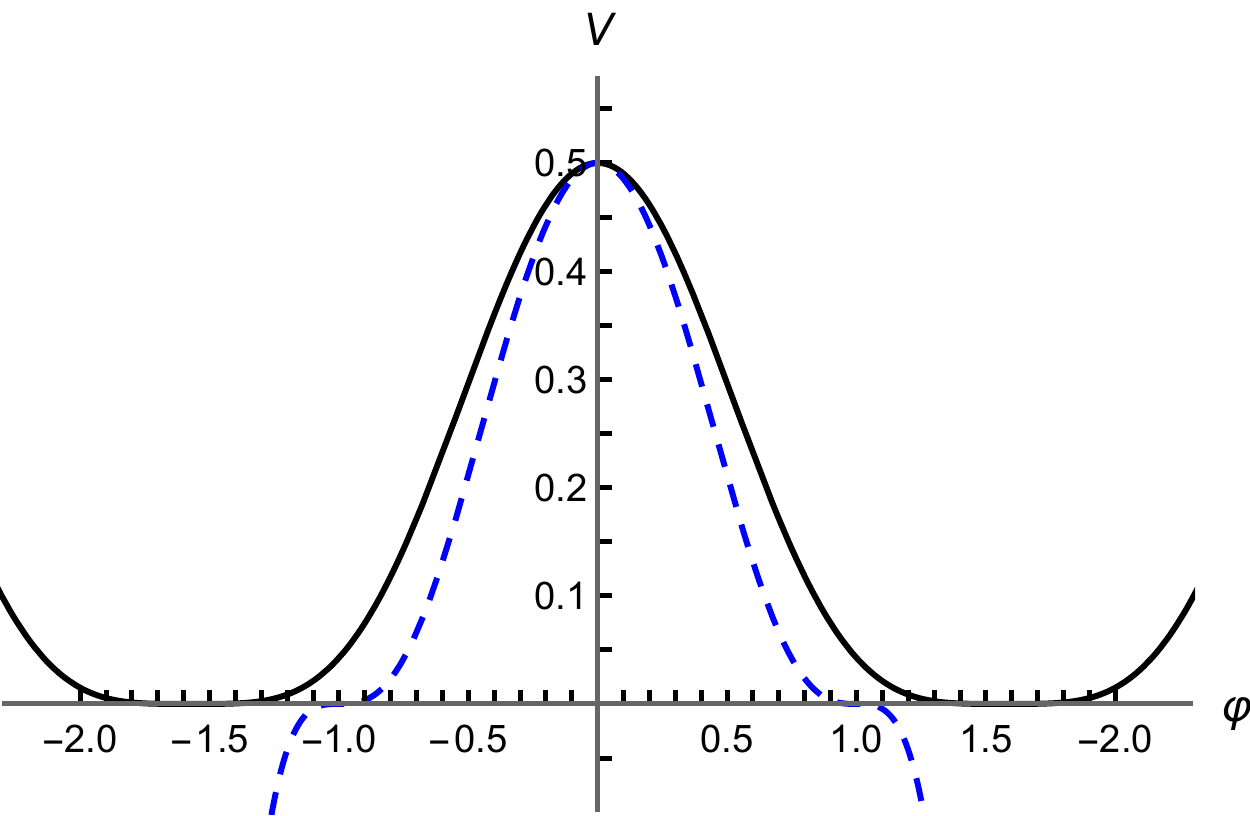}
\label{fig:potentials_example_4}
}
\end{minipage}
\\
\begin{minipage}{0.8\linewidth}
\subfigure[\:Kinks]{
\includegraphics[width=\linewidth]{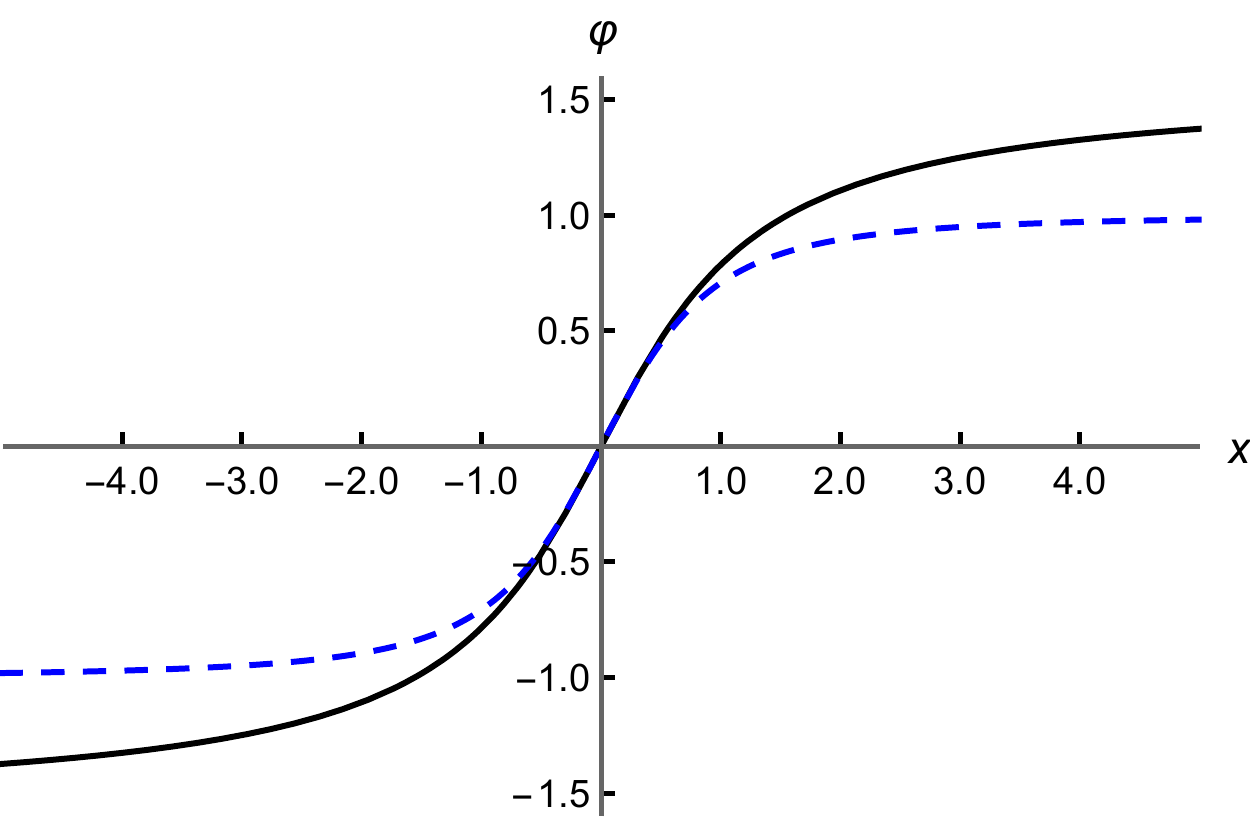}
\label{fig:kinks_example_4}
}
\end{minipage}
\caption{Potentials and kinks of the model \eqref{eq:Fourth_example_potential} (black solid curves) and of the model \eqref{eq:Fourth_example_deformed_potencial} (blue dashed curves).}
\label{fig:potentials_and_kinks_example_4}
\end{figure}

At the same time, for the potential \eqref{eq:Fourth_example_potential}, the kink connecting the vacua $-\pi/2$ and $\pi/2$ can be easily obtained:
\begin{equation}
    \varphi_{\rm K}^{(0)}(x) = \text{arctg}\: x.
\end{equation}
It can be seen that \eqref{eq:Fourth_example_left_asymptotics} accurately describes its asymptotic behavior. On the other hand, the potential of the deformed model has the form
\begin{equation}\label{eq:Fourth_example_deformed_potencial}
    V^{(1)}(\varphi) = \frac{1}{2}\left(1-\varphi^2\right)^3,
\end{equation}
and the corresponding kink
is
\begin{equation}\label{eq:Fourth_example_deformed_kink}
    \varphi_{\rm K}^{(1)}(x) = \frac{x}{\sqrt{1+x^2}}.
\end{equation}
The asymptotic behavior of this kink is exactly described by Eq.~\eqref{eq:Fourth_example_right_asymptotics}. Potentials and kinks of both models are shown in Fig.~\ref{fig:potentials_and_kinks_example_4}.
%
%
%

It is noteworthy that the potential \eqref{eq:Fourth_example_deformed_potencial} is not non-negative for arbitrary $\varphi$. This, however, has no effect (within our approach) on the properties of the kink \eqref{eq:Fourth_example_deformed_kink}, which connects the vacua $-1$ and $1$.

The kink-antikink force in the model \eqref{eq:Fourth_example_potential} decays as fourth power of the inverse separation, in the spirit of Ref.~\cite{Christov.PRL.2019}. At the same time, for the deformed model \eqref{eq:Fourth_example_deformed_potencial} method of Ref.~\cite{Christov.PRL.2019} predicts the sixth power of the inverse separation. The potential \eqref{eq:Fourth_example_deformed_potencial} is not positive-definite outside the interval between the vacua. This means that one should be careful in situations when, in the process of evolution of the configuration, the values of the field may, for some reason, go outside the interval. In some cases, the potential \eqref{eq:Fourth_example_deformed_potencial} should be modified as follows: $V^{(1)}(\varphi)=\displaystyle\frac{1}{2}\left|1-\varphi^2\right|^3$.

The masses of both kinks are $M_{\rm K}^{(0)}=\pi/2$, $M_{\rm K}^{(1)}=3\pi/8$, and $M_{\rm K}^{(1)}<M_{\rm K}^{(0)}$, for the same reason as in the previous example.


\section{Conclusion}
\label{sec:Conclusion}

We have studied the question of how the asymptotic behavior of kink changes when the model is deformed using a bounded function. In the case of a strictly monotonic deforming function with finite derivative, the exponential asymptotics remains exponential with exactly the same coefficient in front of $x$. The power-law asymptotics remains power-law with the same power, but, depending on the derivative of deforming function, the numerical coefficient may change.

This means that in the case of kinks with exponential tails, the force of the kink-antikink interaction retains exponential asymptotic dependence on distance. For kinks with power-law tails, the force retains the same power-law dependence on distance, but it can slightly decrease or increase depending on the value of the derivative of the deforming function in the vacuum point.

Additionally, we have considered the case of a deforming function that has infinite derivative. Namely, we assumed that the derivative of the deforming function goes to infinity in one or both vacua of the deformed model, which are connected by the deformed kink. We found that the exponential asymptotic behavior of the kink remains exponential under this deformation. However, in contrast to the case of finite derivative, the coefficient in front of $x$ now changes. This leads to a faster approach of the field to the vacuum value. The power-law asymptotics of the kink remains power-law. However, the power of the coordinate in the denominator increases, that is, the field of the deformed kink approaches the vacuum value faster.

From the point of view of kink-antikink interaction, this means that for kinks with exponential tails, the interaction force retains the exponential character of the dependence, but for deformed kinks it decreases faster with distance. In the case of power-law tails, in the power-law dependence of the force, the power of distance increases, i.e., again, for deformed kinks, the force decreases faster with distance.

We have also shown that the properties of the deforming function affect how the mass of kink changes when the model is deformed: in the case of $f^\prime(\varphi)>1$, the mass decreases, and in the case of $f^\prime(\varphi)<1$ the mass increases.

Besides that, we have shown how the deformation procedure can be applied to the case of implicit kinks.

As a separate result, we have showed the universality of the asymptotic behavior of the stability potential for any kink with a power-law tail. The stability potential for such kinks asymptotically decreases proportional to the inverse square of the coordinate, approaching zero from above. In the case of a strictly monotonic deforming function with finite derivative, the asymptotics of the stability potential does not change. In the case of infinite derivative of the deforming function, the numerical coefficient in the asymptotics changes.

The asymptotic properties of the kink's stability potential are important for studying the excitation spectrum of the kink. Which, in turn, is of great importance for understanding resonance phenomena in the kink-antikink scattering.

Changes in the asymptotics of kinks for both types of deforming function are illustrated by four examples. As examples, we used well-known field-theoretic models with many physical applications: $\varphi^4$, $\varphi^8$, sine-Gordon, etc. As an example of deforming function with finite derivative we used $f(\varphi)=\sinh\varphi$. The case of a function with infinite derivative is represented by $f(\varphi)=\arcsin\varphi$ in the case of vacua of the original model at points $\pm\pi/2$, which are mapped to vacua of the deformed model $\pm 1$. In all examples we discuss change of the kink-antikink force and kink's mass under deformation.

In conclusion, we would like to formulate some open questions and possible directions for further research.

1. It would be interesting to consider more singular deforming functions that would lead to more drastic changes in the asymptotic behavior of kinks, and hence in the related physics.

2. Deformation procedure can be applied to two-field models \cite{Afonso.PRD.2007}. One could investigate transformational properties of topological defects of such models. Perhaps it might also be interesting (as advised by the reviewer of the manuscript of this paper) to think about generalizations of the deformation procedure to kink solutions of more complex models \cite{Takahashi.PRL.2013}, but this could be a very hard problem.

3. The stability potential of a kink with two power-law tails is volcano-like with the only zero mode in the spectrum, see, e.g., \cite[Sec.~5]{Gani.JPCS.2020.no-go}. Nevertheless, there could be quasi-discrete levels (quasi-normal modes) in the continuous part of the spectrum. Searching for such levels could become a subject of future study. We would like to study this issue and, if successful, hope to report our results in the near future.


\section*{Acknowledgments}

We would like to thank the Reviewer for a number of comments that have helped us to improve the manuscript. V.A.G.\ is also grateful to Prof.\ Nicholas Manton for helpful discussions.

V.A.G.\ acknowledges the support of the Russian Foundation for Basic Research under Grant No.\ 19-02-00971. This work was also supported by the MEPhI Program ``Priority-2030''.


\end{document}